\documentclass{aa}  
\usepackage{natbib}
\usepackage{graphicx}
\usepackage{aalongtable}
\usepackage{txfonts}
\begin{document}
   \title{Ensemble X-ray variability of active galactic nuclei from serendipitous source catalogues\thanks{Tables 1 and 2 are available in electronic form at {\tt http://www.aanda.org}}}

   \author{F. Vagnetti \inst{1}\and S. Turriziani \inst{1}\fnmsep\thanks{Visitor at ASI Science Data Center, c/o ESRIN, via G. Galilei, I-00044 Frascati, Italy} \and D. Trevese \inst{2}}
   
   \institute
       {Dipartimento di Fisica, Universit\`a di Roma ``Tor Vergata'', Via della Ricerca Scientifica 1, I-00133, Roma, Italy\\ \email{fausto.vagnetti@roma2.infn.it}
      \and
       Dipartimento di Fisica, Universit\`a di Roma ``La Sapienza'', Piazzale Aldo Moro 2, I-00185 Roma, Italy}   
   \date{Received 12 September 2011 / Accepted 20 October 2011}
   
  \abstract
   {The X-ray variability of the active galactic nuclei (AGN) has been most often investigated with studies of individual, nearby sources, and only a few ensemble analyses have been applied to large samples in wide ranges of luminosity and redshift.}
   {We aim to determine the ensemble variability properties of two serendipitously selected AGN samples extracted from the catalogues of XMM-Newton and Swift, with redshift between $\sim 0.2$ and $\sim 4.5$, and X-ray luminosities, in the 0.5-4.5 keV band, between $\sim 10^{43}$ erg/s and $\sim 10^{46}$ erg/s.}
   {We used the structure function (SF), which operates in the time domain, and allows for an ensemble analysis even when only a few observations are available for individual sources and the power spectral density (PSD) cannot be derived. The SF is also more appropriate than fractional variability and excess variance, because these parameters are biased by the duration of the monitoring time interval in the rest-frame, and therefore by cosmological time dilation.}
   {We find statistically consistent results for the two samples, with the SF described by a power law of the time lag, approximately as $S\negthinspace F\propto  \tau^{0.1}$. We do not find evidence of the break in the SF, at variance with the case of lower luminosity AGNs. We confirm a strong anti-correlation of the variability with X-ray luminosity, accompanied by a change of the slope of the SF. We find evidence in support of a weak, intrinsic, average increase of X-ray variability with redshift.}
   {The change of amplitude and slope of the SF with X-ray luminosity provides new constraints on both single oscillator models and multiple subunit models of variability.}

   \authorrunning{F.Vagnetti et al.}
\titlerunning{X-ray variability of AGNs}

   \maketitle

\section{Introduction}
Active galactic nuclei (AGN) typically show flux variability in all wavebands and on different timescales from minutes to years. This behaviour has been widely used to constrain the size and location of the emission regions and to obtain information on the emission mechanisms as well as the processes that cause the variability itself.

In addition to the study of individual light curves, ensemble properties of statistical AGN samples have been investigated in the optical/UV band through the use of the structure function (SF) \citep[e.g.][]{trev94,cris96,vand04}, and in the X-rays through the analysis of the fractional variability  \citep{alma00,mann02}.

Optical variability has been found to increase with decreasing luminosity \citep[e.g.][]{cris96}, and with increasing redshift \citep{gial91}. The average increase with redshift of the amplitude of variability  can be explained by the fact that high-redshift sources are observed at a higher rest-frame frequency, where they are more variable \citep{di-c96}. The stronger variability at higher frequency, in turn, is caused by a hardening of the spectral energy distribution (SED) in the brighter phase, as shown by ensemble analyses of multiband optical photometry of quasar (QSO) samples \citep{trev01,trev02}. More recently, \citet{vand04} applied an ensemble analysis to a large sample of $\sim 25,000$ QSOs observed at two epochs only with the Sloan Digital Sky Survey (SDSS). The authors analysed variability as a function of intrinsic luminosity, redshift, rest-frame frequency and time lag between the observations, proposing a weak, intrinsic increase of variability with redshift, in addition to the amount previously explained by the stronger variability at higher rest-frame frequency \citep{di-c96}, although additional analyses have not confirmed this increase \citep[e.g.][]{macl10}.

In the X-ray domain, variability occurs on shorter time scales than in any other band, and is thought to come from a hot corona close to the central black hole (BH). Most investigations concern the light curves of individual nearby Seyfert 1 AGNs \citep[e.g.][]{uttl02,uttl05a}. It has been found that low-luminosity AGNs are generally more variable than higher luminosity ones \citep[e.g.][]{barr86,lawr93,gree93,nand97}, and that the variability amplitude is higher on long time scales than on short time scales \citep[e.g.][]{mark04}. In addition, it has been suggested that variability also increases with redshift \citep{paol04}.

Proposed variability models include a single coherent oscillator \citep[e.g.][]{alma00}, a superposition of individual flares or spots \citep[e.g.][]{leht89,abra91,czer04}, variable absorption and/or reflection \citep[e.g.][]{abra00,mini04,chev06}.

The relation between X-ray and optical/UV variability may be due to either i) Compton up-scattering in the hot corona of optical photons emitted by the disk \citep{haar91}, or to ii) a reprocessing of X-rays into thermal optical emission by means of irradiation and heating of the accretion disk \citep{coll91}. In the first case, variations in the optical/UV flux would lead to X-ray variations, and vice versa in the latter case. Cross-correlation analyses of well-sampled X-ray and optical/UV light curves allow us to constrain models for the cause of the variability. The main results obtained so far indicate a cross-correlation between X-ray and UV/optical variations on the timescale of days, and in some cases delays between the two bands have been measured, with both X-rays lagging the UV \citep[e.g.][]{mars08,doro09}, and vice versa \citep[e.g.][]{shem01,arev09}.

Even more insight into the relation between X-ray and optical/UV variability is given by the analysis of the X-ray/UV ratio and its variability. \citet{vagn10} have shown that variability of $\alpha_{ox}$\footnote{$\alpha_{ox}\equiv\log(L_{\rm 2\,keV}/L_{\rm 2500\AA})/\log(\nu_{\rm 2\,keV}/\nu_{\rm 2500\AA})$} increases as a function of time-lag for a sample of serendipitously selected AGNs with simultaneous X-ray and UV measurements. This contributes part of the observed dispersion in the $\alpha_{ox}$-$L_{UV}$ anti-correlation, while another contribution is given by intrinsic differences among the average values of each AGN.

In the present paper, we present for the first time an ensemble structure function analysis of the variability of AGNs in the X-ray band. We adopt two sets of multi-epoch X-ray measurements extracted from the serendipitous source catalogues of XMM-Newton \citep{wats09} and Swift \citep{pucc11}.

The paper is organised as follows. Section 2 describes the data extracted from the archival catalogues. Section 3 describes the computation of the structure functions, and discusses their shapes, their dependence on black hole mass and bolometric luminosity, as well as on X-ray luminosity and redshift. In Sect. 4 we discuss and summarise the results.

Throughout the paper, we adopt the cosmology $H_0= 70 {\rm \,km\,s^{-1}\,Mpc^{-1}}$, $\Omega_m=0.3$, and $\Omega_\Lambda=0.7$.

\section{The data}
\subsection{XMM-Newton}
The XMM-Newton Serendipitous Source Catalogue (XMMSSC) \citep{wats09} is a comprehensive catalogue of serendipitous X-ray sources from the XMM-Newton observatory. The version presently available is 2XMMi-DR3, the latest incremental update of the second version of the catalogue\footnote{\tt http://xmmssc-www.star.le.ac.uk/Catalogue/}.  It contains source detections drawn from 4953 XMM-Newton EPIC observations made between 2000 February 3 and 2008 October 08; all datasets were publicly available by 2009 October 31, but not all public observations are included in this catalogue. The total area of the catalogue fields is $\sim$814 deg$^2$, but taking account of the substantial overlaps between observations, the net sky area covered independently is $\sim$504 deg$^2$. The 2XMMi-DR3 catalogue contains 353191 detections (above the processing likelihood threshold of 6), related to 262902 unique X-ray sources, therefore a significant number of sources (41979) have more than one record within the catalogue. 

We used the TOPCAT\footnote{\tt http://www.star.bris.ac.uk/$\sim$mbt/topcat/} software to extract the sources with repeated X-ray observations from the 2XMMi-DR3 catalogue and cross-correlated this list with the DR7 edition of the SDSS Quasar Catalogue \citep{schn10} to obtain redshifts and spectral classifications of the sources. We used a maximum distance of 1.5 arcsec, corresponding to the uncertainty in the X-ray position, resulting in 412 quasars that were observed from 2 to 25 epochs each for a total of 1376 observations. We refer to these sources as the {\it XMM-Newton sample}, and report them in Table 1, where
Col. 1 corresponds to the source serial number;
Col. 2 gives the source name;
Col. 3 the redshift;
Col. 4 the number of observation epochs for the source;
Col. 5 the average log of the X-ray flux in the observed 0.5-4.5 keV band, in erg cm$^{-2}$ s$^{-1}$;
Col. 6 the average log of the X-ray luminosity in the 0.5-4.5 keV band, in erg s$^{-1}$, computed with a  photon index $\Gamma=1.8$;
and Cols. 7 and 8 the log of the minimum and maximum lag between any two epochs of the light curve in the rest-frame of the source in days.

\begin{figure}
\centering
\resizebox{\hsize}{!}{\includegraphics{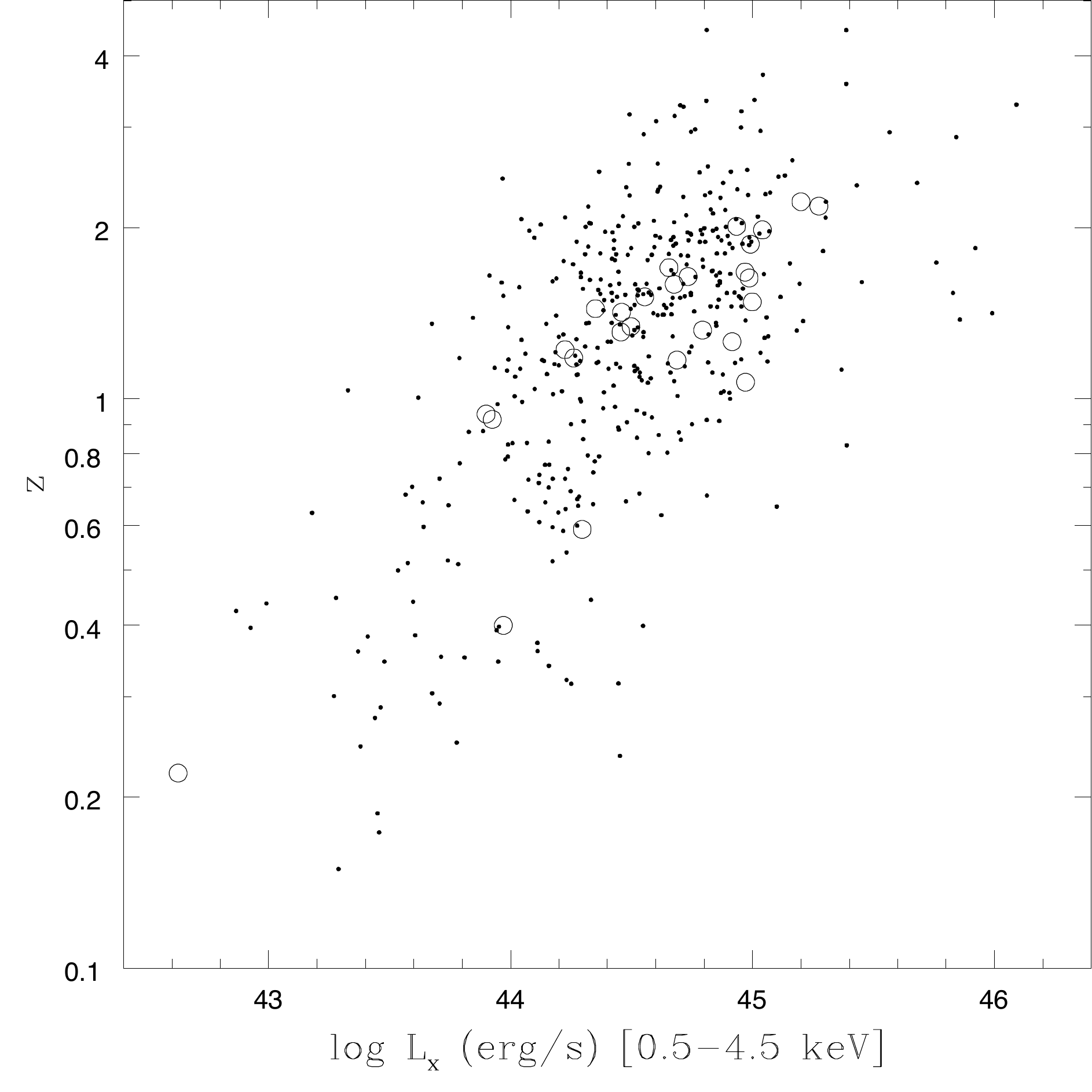}}
\caption{Distribution of the sources in the $L_X$-$z$ plane. Dots: XMM-Newton sample; circles: Swift sample.}
\end{figure}

The sources are shown in the $L_X$-$z$ plane in Fig. 1 together with the sources of the Swift sample (Sect. 2.2). Here and throughout we adopted the same X-ray band 0.5-4.5 keV for the two samples. For XMMSSC, the flux was directly extracted from the EP9 band of the catalogue, while for the Swift sample the flux was computed from the Swift band 0.3-10 keV, adopting a photon index $\Gamma=1.8$.

Typical monitoring times range from months to few years in the rest-frame. Some of the best sampled light curves with 10 or more epochs are shown in Fig. 2, with times in rest-frame days, counted from the initial epoch of each light curve.

\begin{figure}
\centering
\resizebox{\hsize}{!}{\includegraphics{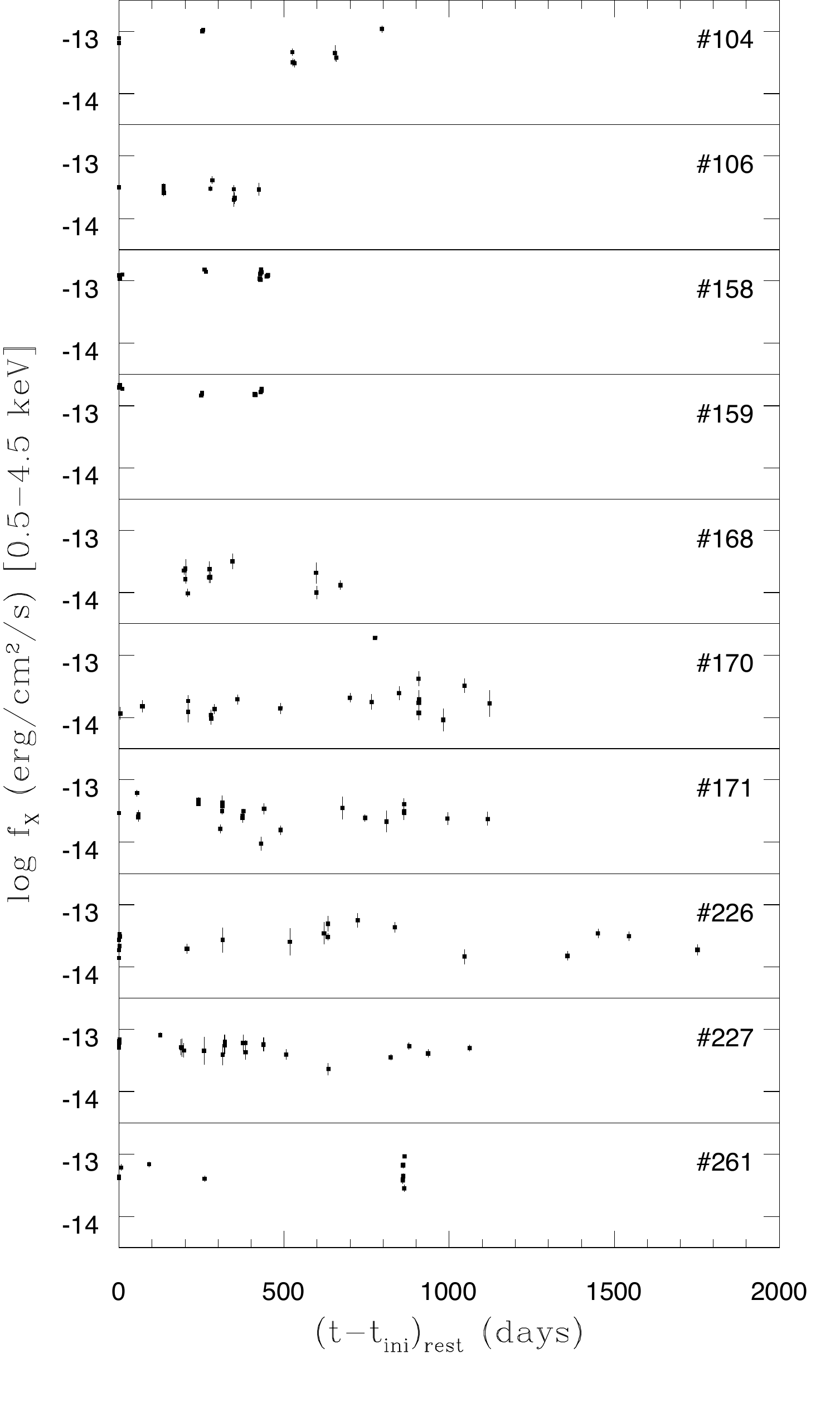}}
\caption{Some of the best-sampled light curves from the XMM-Newton sample. Times are counted from the initial epoch of each light curve in the rest-frame. Errors, proportional to the inverse square root of the photon count at each epoch, are displayed as 3-$\sigma$ values. Errors are discussed in more detail in Section 3. Source numbers from column 1 of Table 1 are indicated.}
\end{figure}

\subsection{Swift}
In the context of serendipitous surveys, the Swift satellite provides a unique capability. Although this space observatory is designed to discover gamma-ray bursts (GRB) \citep{gehr04}, it is possible to use individual pointed observations of each GRB to build a large sample of deep X-ray images by stacking the individual exposures. To this purpose, \citet{pucc11} considered all Swift GRB observations from January 2005 to December 2008, with a total exposure time in the X-ray Telescope (XRT) longer than 10 ksec. These authors also analysed the XRT 0.5 Msec observation of the Chandra Deep Field South (CDFS) sky region. This set of observations is called the Swift Serendipitous Survey in deep XRT GRB Fields (S3XGF). These 374 images make up an unbiased X-ray survey because GRBs explode at random positions in the sky, and \citet{pucc11} used them to define a well-suited statistical sample of X-ray point sources. The total exposure time of the survey is 36.8 Msec, with $\sim32\%$ of the fields with more than 100 ksec exposure time, and $\sim 28\%$ with exposure time in the range 50-100 ksec. The survey covers a total area of $\sim$ 32.55 deg$^2$.

We used the preliminary version S3XGF catalogue, comprising GRB fields observed from January 2005 to June 2007, and cross-correlated it with the DR7 edition of the SDSS Quasar Catalogue \citep{schn10} to obtain redshifts and spectral classifications.

We found 27 confirmed quasars with sufficient sampling (at least 100 photons in the light curve) to be used in the following SF analysis. These sources, to which we will refer as the {\it Swift sample}, are reported in Table 2, where
Col. 1 corresponds to the source serial number;
Col. 2 to the source name;
Col. 3 gives the redshift;
Col. 4 the number of time bins into which we divide the light curve according to the procedure described in the following;
Col. 5 the average log of the X-ray flux in the band 0.5-4.5 keV, in erg cm$^{-2}$ s$^{-1}$;
Col. 6 the average log of the X-ray luminosity in the band 0.5-4.5 keV, in erg s$^{-1}$;
Col. 7 the GRB field where the source was observed.

The light curve files extracted from the Swift archive contain sequences of time intervals $\Delta t_i$ between $t_{start,i}$ and $t_{stop,i}$, in which the telescope was observing, with $n_i$ the number of photons detected in each interval. We binned the light curves using a bin size $\Delta t_{bin}=5\times 10^4$ s, which is a good compromise to obtain an average number of photons/bin $\ga 10$ and a number of useful bins (i.e., bins with non-zero number of photons) in the light curve $\ga 10$. There is a negligible number of bins with zero photons, however. We assigned an average time $t_j$ to each bin $j$  weighted by the number of photons detected in the intervals (or fractions of intervals) $\Delta t_i$ overlapped with the bin: $t_j=\sum n_i t_i/\sum n_i$, where $t_i=(t_{start,i}+t_{stop,i})/2$. 

\begin{figure}
\centering
\resizebox{\hsize}{!}{\includegraphics{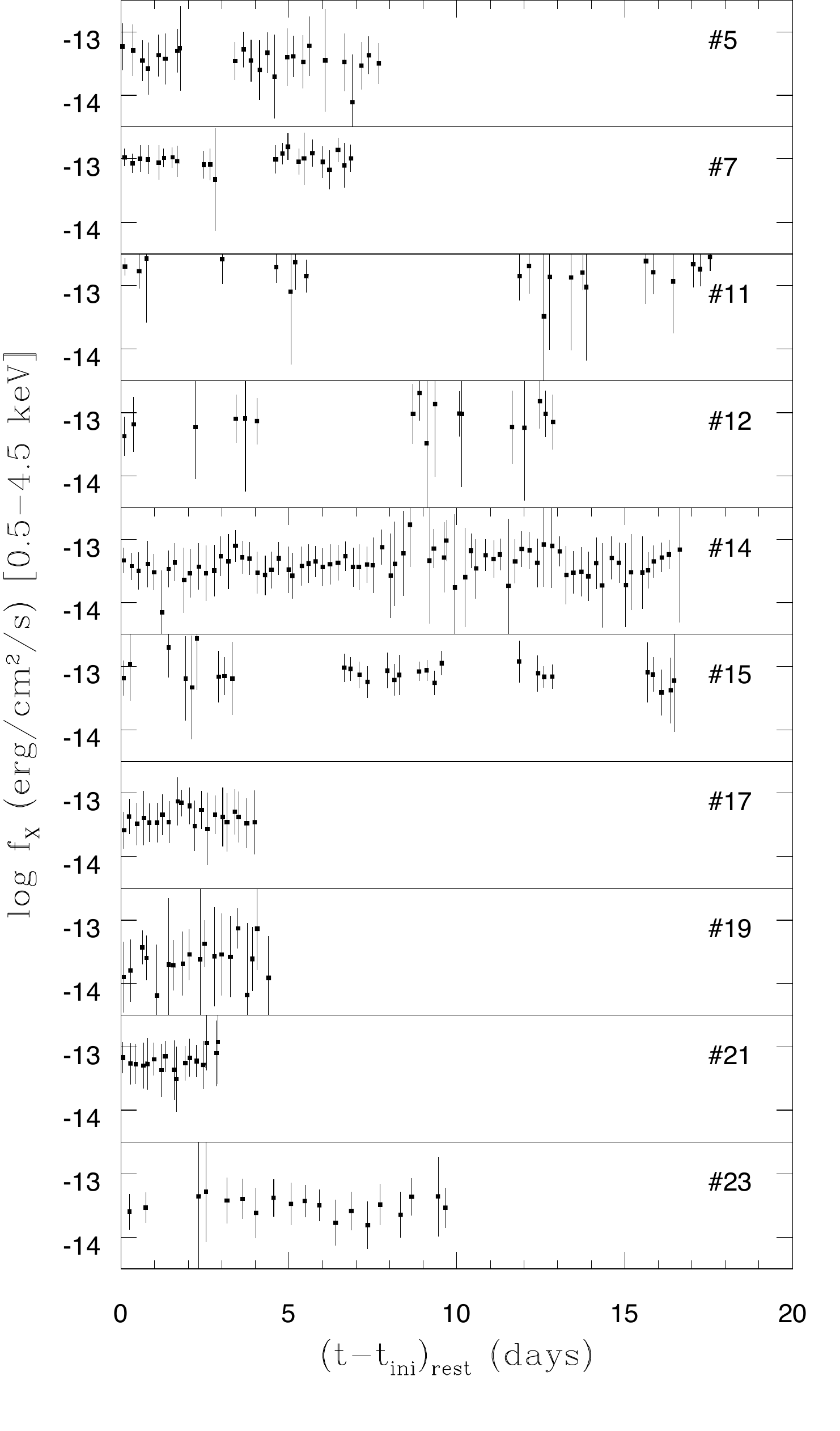}}
\caption{Some of the best-sampled light curves from the Swift sample. Times are counted from the initial epoch of each light curve, in the rest-frame. Errors proportional to the inverse square root of the photon count at each epoch are displayed as 3-$\sigma$ values. Errors are discussed in more detail in Section 3. Source numbers from column 1 of Table 2 are indicated.}
\end{figure}

Some of the best-sampled light curves are shown in Fig. 3 with times in rest-frame days counted from the initial epoch of each light curve. Typical monitoring times range from some days to a few weeks in the rest-frame, and are therefore complementary to the time scales sampled by XMM-Newton.

The distribution of the Swift sample in the $L_X$-$z$ plane is shown in Fig. 1 together with the XMM-Newton sample.

\section{The structure function}
The structure function (SF) has the great advantage of working in the time domain, which allows for an ensemble analysis even for extremely poor sampling of individual objects, when the armonic content is completely lost. In this case, the structure function is to be preferred over power spectral density (PSD) analysis \citep[e.g.][]{hugh92,coll01,favr05}. The SF was first introduced by \citet{simo85}, and has since been used in various bands, including radio \citep[e.g.][]{hugh92}, optical \citep[e.g.][]{trev94,kawa98,de-v03,baue09}, and X-ray \citep[e.g.][]{fior98,brin01,glio01,iyom01,zhan02}.

The SF provides a measure of the mean deviation for data points separated by a time lag $\tau$, and is defined in various ways in the literature. A variant in the definition concerns the use of the average square difference \citep[e.g.][]{simo85,hugh92} or the average of  the absolute values of the differences \citep{di-c96}. Another variant concerns the use of magnitudes or fluxes: while in the optical the SF is usually defined in terms of magnitude differences, in the X-rays and in the radio band the SF  is most often defined in terms of flux differences, although there are exceptions, e.g. \citet{fior98} introduced X-ray magnitudes and their differences. 

For an analogy with the optical, we used the logarithm of the flux instead of the flux itself, and defined the SF with the following formula:

\begin{equation}
S\negthinspace F(\tau)\equiv\sqrt{{\pi\over 2}\langle|\log f_X(t+\tau)-\log f_X(t)|\rangle^2-\sigma_n^2}\quad .
\end{equation}

\noindent
Here, the average of the absolute value of the difference is used, as in \citet{di-c96};  $\sigma_n$ is the contribution of the photometric noise to the observed variations. $f_X(t)$ and $f_X(t+\tau)$ are two measures of the flux $f_X$ in a given X-ray band at two epochs differing by the lag $\tau$. The factor $\pi/2$ normalises SF to the rms value in the case of a Gaussian distribution. The X-ray band adopted in this paper is 0.5-4.5 keV, and the lag $\tau$ is computed in the rest frame:

\begin{equation}
\tau_{rest}=\tau_{obs}/(1+z) \quad .
\end{equation}

While a definition in terms of flux differences could also be used for studies of individual sources, our definition with logarithmic differences, Eq. (1), is certainly preferable for an ensemble analysis, otherwise the contribution of faint sources would be negligible compared to that of brighter ones. 

\begin{figure}
\centering
\resizebox{\hsize}{!}{\includegraphics{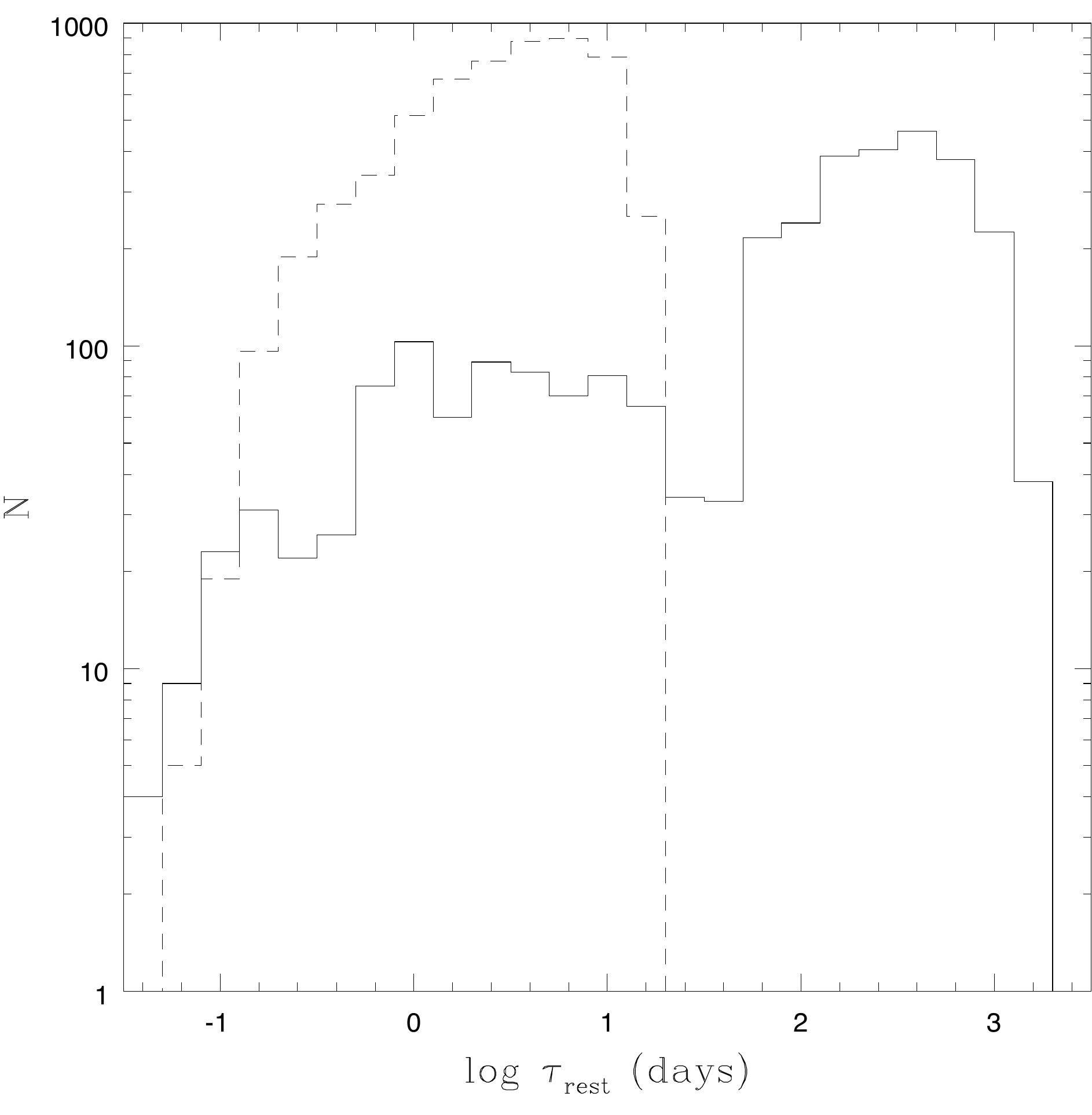}}
\caption{Histogram of the rest-frame time lags contributing to the structure functions. 
Continuous histogram: XMM-Newton sample; dashed histogram: Swift sample.}
\end{figure}

We now computed and compared the structure functions for the two samples. While the XMM-Newton sample is much larger than the Swift sample (412 sources vs 27 sources), the number of epochs is very small for most XMM-Newton sources (338/412=82\% of the sources having less than 5 epochs), while the Swift light curves (with the adopted binning, see Sect 2.2) are better sampled, 21/27=78\% of the sources having 10 or more bins (or ``epochs"), with a mean number $\sim$ 16. So the contributions of the two samples to the respective SFs are comparable in number, although different in the time scales sampled. The light-curve of the k-th source, with $N_k$ epochs, contributes $N_k(N_k-1)/2$ points to $S\negthinspace F(\tau_{rest})$, for all the time lags $\tau_{rest,ij}=|t_i-t_j|/(1+z)$, where $t_i$ and $t_j$ are two epochs in the observer frame.

This can be seen in Figure 4, where the histograms of the rest-frame time lags are shown for the two samples, with bins of $\Delta\log\tau=0.2$: hundreds of points contribute the most populated bins of each sample, which are days-weeks for the Swift sample and months-years for the XMM-Newton sample. The latter contributes also non-negligibly in the days-weeks range, with several tens of points.

In Figs. 5 and 6 we show the structure functions computed with Eq. (1) for the XMM-Newton and Swift samples, respectively. 

\begin{figure}
\centering
\resizebox{\hsize}{!}{\includegraphics{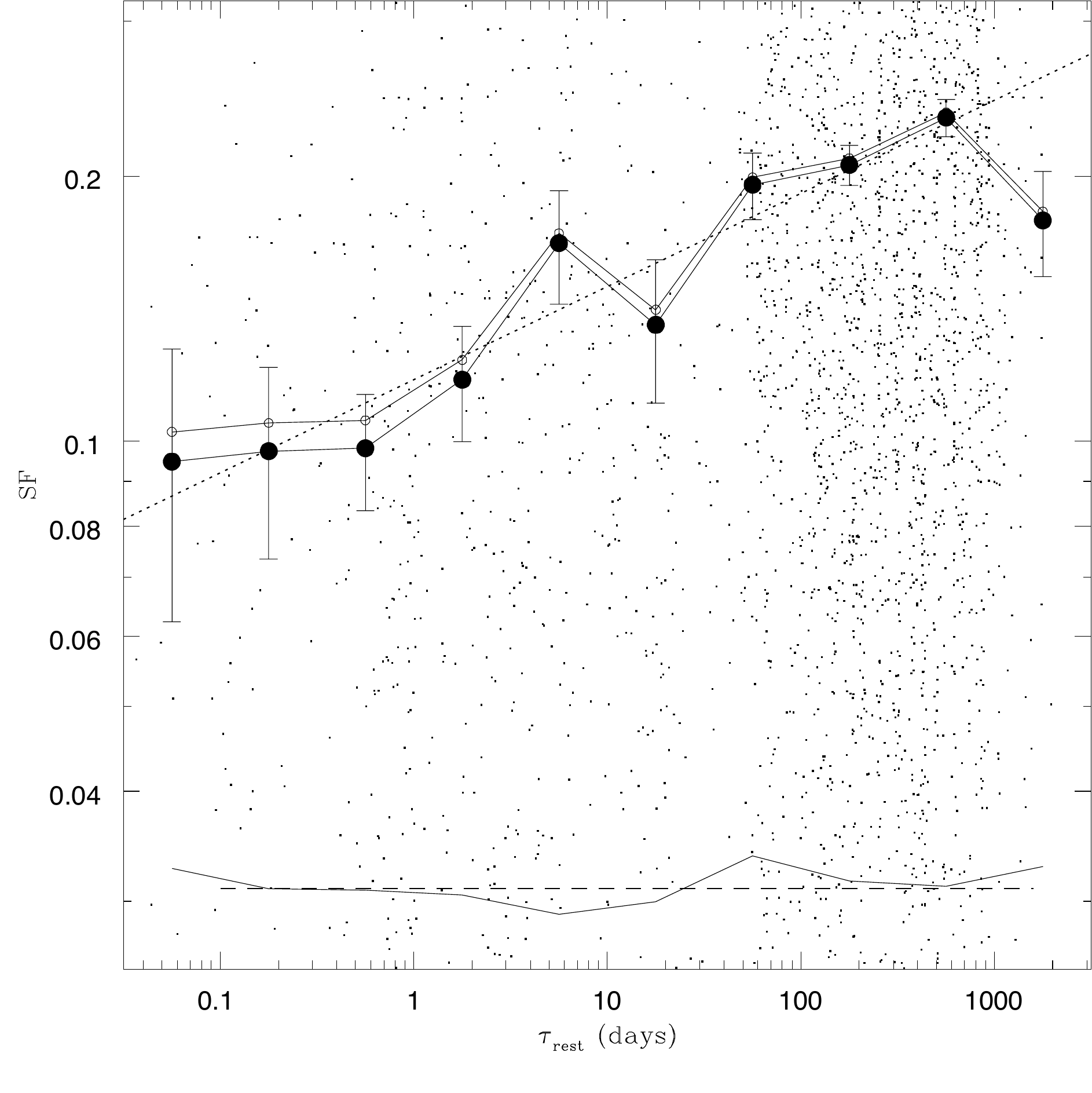}}
\caption{Structure function for the XMM-Newton sample in bins of $\Delta\log\tau=0.5$. The small empty circles and the continuous line connecting them show the uncorrected SF (i.e., neglecting $\sigma_n$ in Eq. (1)). The larger filled circles and the line connecting them, show the SF corrected for the noise. The continuous line without data points indicates the average value of the noise in each bin, and the dashed, horizontal line is its weighted average, according to the number of points in each bin, adopted in Eq. (1). The dotted line is a weighted least-squares fit to the data of the bins. The small dots are the contributions from pairs of individual measurements at times differing by $\tau$.}
\end{figure}

\begin{figure}
\centering
\resizebox{\hsize}{!}{\includegraphics{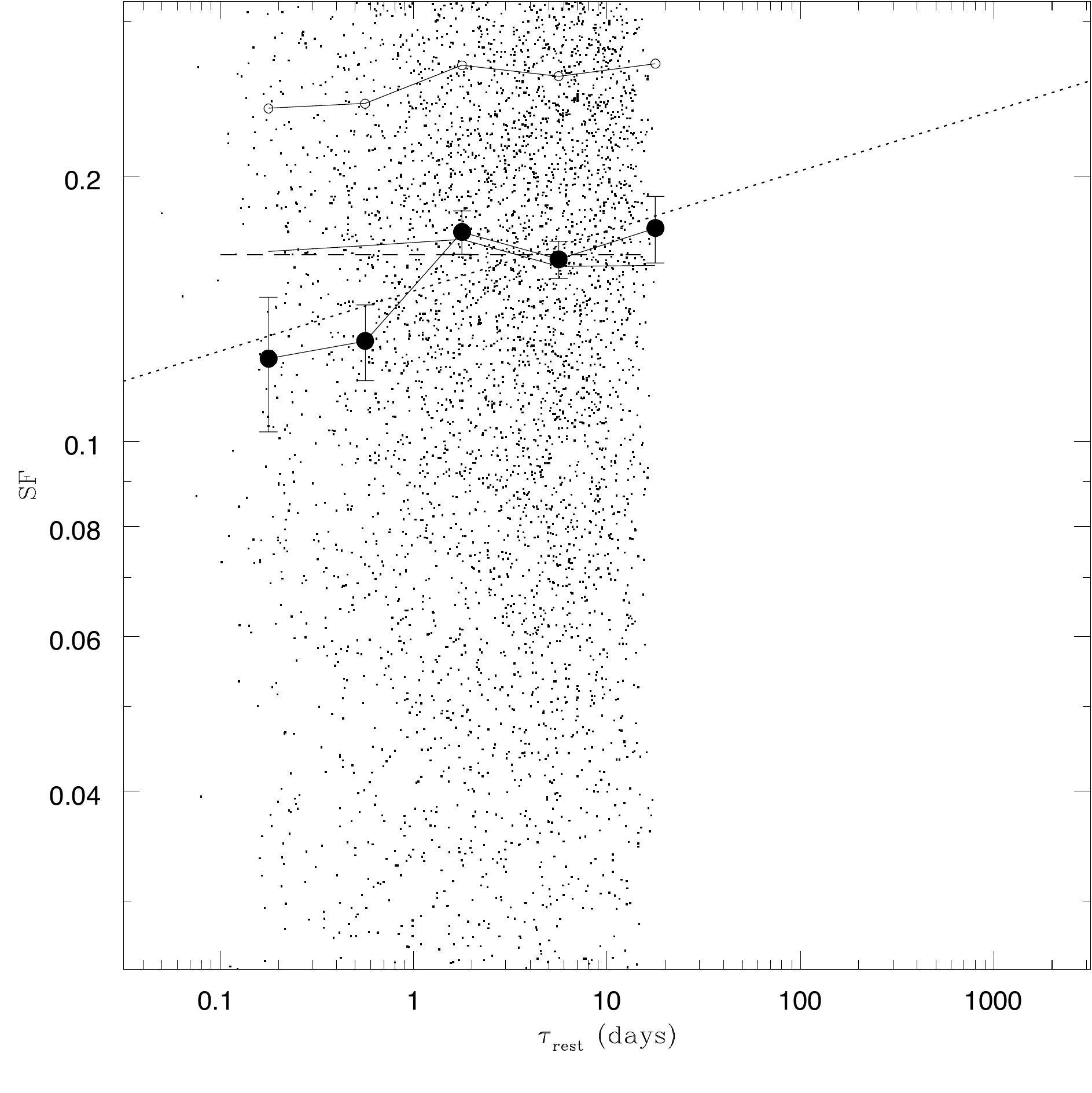}}
\caption{Structure function for the Swift sample, in bins of $\Delta\log\tau=0.5$. The small empty circles and the continuous line connecting them show the uncorrected SF (i.e., neglecting $\sigma_n$ in Eq. (1)). The larger, filled circles and the line connecting them show the SF corrected for the noise. The continuous line without data points indicates the average value of the noise in each bin, and the dashed, horizontal line is their weighted average, according to the number of points in each bin, adopted in Eq. (1). The dotted line is a weighted least-squares fit to the data of the bins. The small dots are the contributions from pairs of individual measurements at times differing by $\tau$.}
\end{figure}

To estimate the photometric noise $\sigma_n$ in Eq. (1), we evaluated its contribution in each bin with the following considerations. The quadratic contribution of the noise to the SF is

\begin{eqnarray}
\sigma_n^2=2\langle(\delta\log f_X)^2\rangle\approx 2(\log e)^2\left\langle\left({\delta f_X\over f_X}\right)^2\right\rangle = \nonumber \\
= 2\cdot 0.434^2\left\langle{1\over N}\right\rangle=0.377{\sum{1\over N_k}\over N_p} ~,
\end{eqnarray}

\noindent
where $\delta f_X$ are the flux variations caused by noise alone (excluding source variability), and we assume $\delta f_X/f_X= 1/\sqrt{N}$, $N$ being the number of counted photons at a given epoch, and its reciprocal is mediated in any given bin of the SF among the $N_p$ points, which are contributed by the various light-curves;  $N_k$ is the average photon count per epoch of the k-th light-curve; the factor 2 is due to the contribution of 2 independent measurements to each flux variation. The values obtained in each bin are connected and shown in Figs. 5 and 6 as thin, continuous lines, while their average values, weighted with the numbers of points in each bin, are shown as dashed lines.

The average values $\sigma_n=0.031$ (XMM-Newton sample) and $\sigma_n=0.163$ (Swift sample) were then inserted in Eq. (1) to compute the SF, which is shown in Figs. 5 and 6, both with and without noise subtraction. The noise so estimated is almost negligible for the XMM-Newton sample, and quite high for the Swift sample. This is mainly because of the smaller effective area of Swift, and also because of the longer exposures of the XMM-Newton observations, which are typically several tens of ks per epoch, while for Swift the light-curves are binned in intervals of 50 ks, with effective exposures within a small fraction of the bin, around 10 ks.

Although the two SFs appear different before noise subtraction, their slopes and amplitudes agree quite well after correction. We stress that noise subtraction is not parametrical, but consistently derived by the photon counts. The fits shown in Figures 5 and 6 are least squares of the bin representative points, weighted with the number of individual points in each bin, $\log S\negthinspace F= a + b \log \tau_{rest}$, or

\begin{equation}
S\negthinspace F \propto \tau_{rest}^b
\end{equation}

\noindent
with consistent slopes, $b=0.10\pm 0.01$ for the XMM-Newton sample, and $b=0.07\pm 0.04$ for the Swift sample.

\subsection{Relation with the PSD}
X-ray variability of individual sources is usually analysed in terms of the PSD. This has been often described by a power-law, $P(f)\propto f^{-\alpha},\,\alpha \sim 1.5$, \citep[e.g.][]{lawr93}. However, one or two breaks in the PSD of nearby AGNs have also been detected \citep[e.g.][]{mark03, onei05}, and the PSD has been found to have a power-law exponent $\alpha\approx 2$ for $f>f_{HFB}$, $\alpha\approx 1$ for $f_{LFB}<f<f_{HFB}$, and in some cases $\alpha\approx 0$ for $f<f_{LFB}$. In turn, the high-frequency break has been found to be related to the mass of the central BH \citep[e.g.][]{papa04}. 

An SF with the form of a single power-law as in Eq. (4) is equivalent to a single power-law PSD if the frequency range extends from 0 to $\infty$. Then a simple relation between the exponents holds \citep[e.g.][]{kawa98,baue09,emma10}:

\begin{equation}
\alpha=1+2b~ .
\end{equation}

The slope of our SF, $b\la 0.1$, would then correspond to a PSD exponent $\alpha\la 1.2$, slightly flatter than the reference value $\alpha\sim 1.5$ \citep{lawr93}.

However, Eq. (5) does not straightforwardly apply when the PSD contains a break. \citet{emma10} produced 2000 artificial light-curves with a PSD shaped as a broken power-law with a break at a given value $f_B$, and estimate the corresponding SFs. Figures 10 and 11 of \citet{emma10} show that SFs also display a break whose distribution peaks around $\tau_B\sim 1/f_B$, but the SF slopes before and after this break do not agree with the relation of Eq. (5). In particular, the SF appears flatter than Eq. (5) below the break, and steeper above the break, resulting in less bending.

To analyse the relation between the shapes of PSD and SF, we evaluated the SF numerically via fast Fourier transform (FFT) techniques according to the relation $S\negthinspace F(\tau)=2\int_0^\infty(1-\cos(2\pi f\tau)P(f)df$ \citep[e.g.][]{emma10}, for a PSD shaped as a broken power-law. The result, shown in Fig. 7 for input PSD spectral indexes $\alpha_1=1.2$,  $\alpha_2=2$, is an SF shaped approximately as a broken power-law, but with a slope changing gradually and with less bending, which confirms the result by \citet{emma10}. 

The above results suggest that we should expect some evidence of a break  in the SF of AGNs with a typical broken power-law PSD.

\begin{figure}
\centering
\resizebox{\hsize}{!}{\includegraphics{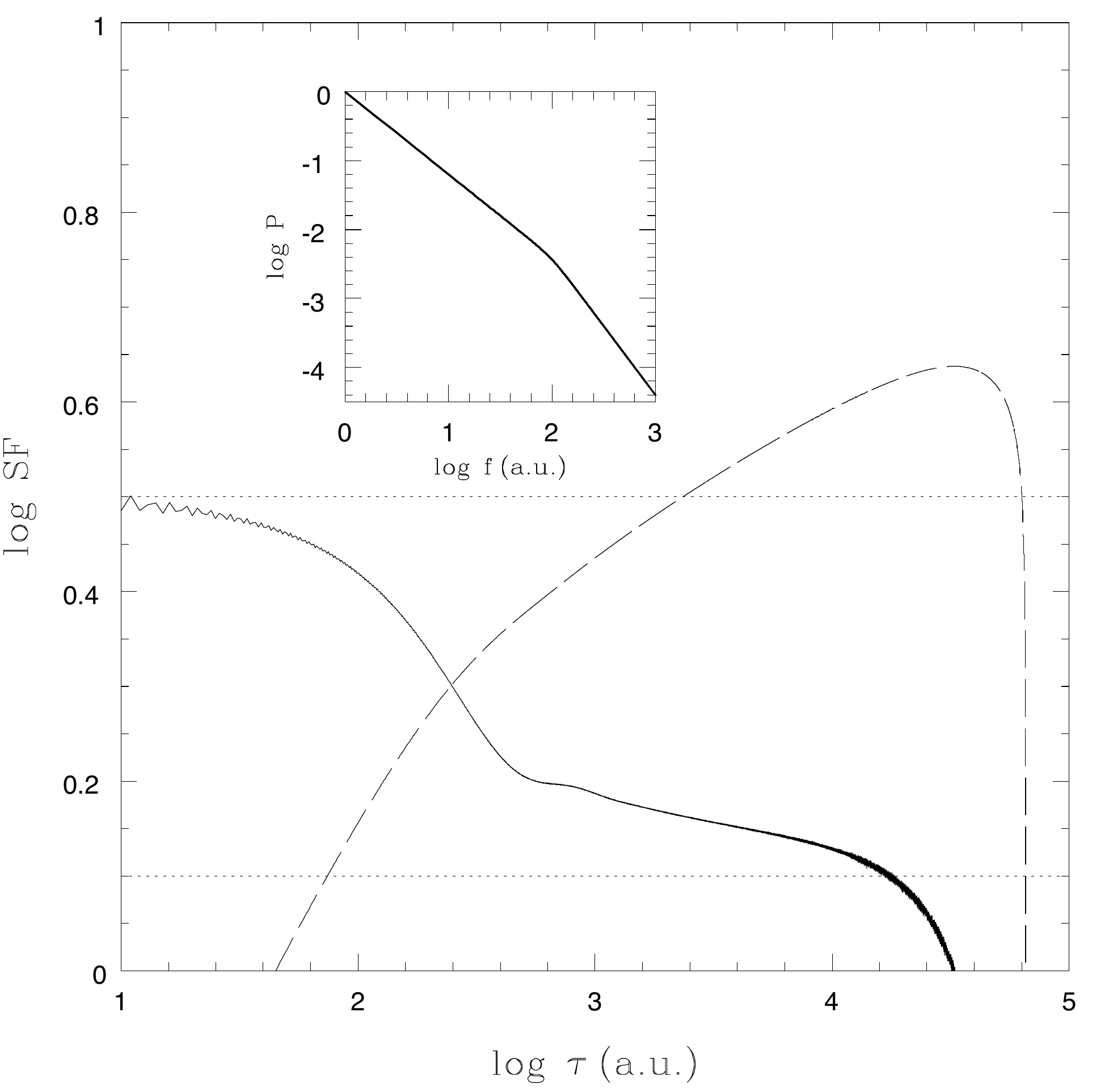}}
\caption{SF (dashed line) and its slope (continuous line) computed from a PSD shaped as a broken power-law, with $\alpha_1=1.2$ and $\alpha_2=2$, shown in the inset. The dotted lines indicate the values of the power-law exponent $b$, expected from Eq. (5) for the single power-law case, $b_1=0.1$, $b_2=0.5$. The cut-off in the SF at long $\tau$ is caused by the finite number of Fourier frequencies used in the FFT calculation. Time lags and Fourier frequencies are in arbitrary units.} 
\end{figure}

\subsection{Dependence on mass and bolometric luminosity}
\citet{mcha06} proposed that the high-frequency break is related not only to the black hole mass, $M_{BH}$, but also to the accretion rate in units of its Eddington value, $\dot{m}_E\approx L_{bol}/L_{Edd}$, and found the following relation

\begin{equation}
\log\tau_{break}{\rm (days)} = 2.1\log M_6 -0.98\log L_{44} -2.32\quad,
\end{equation}

\noindent
where we abbreviate $M_6=M_{BH}/10^6M_\odot$ and $L_{44}=L_{bol}/10^{44}$ erg/s.

However, while for the light curves of individual objects the relation between SF and PSD is relatively simple, for an ensemble SF the different positions of the breaks should combine in the ensemble SF, possibly smoothing the result, depending on how the variability amplitude changes with $M_{BH}$ and/or $L_{bol}$.

To find any break in the SF, we segregated our XMM-Newton sample according to $M_{BH}$ and $L_{bol}$ values. Estimates of the masses and bolometric luminosities were extracted from the catalogue of quasar properties by \citet{shen11}.  We show in Figure 8 the distribution of $\sim 100,000$ AGNs from that catalogue in the plane $M_{BH}$-$L_{bol}$, as well as the same distribution for our XMM-Newton sample, which appears quite similar, despite its smaller population (412 sources). We also show in the same figure some low-luminosity AGNs from \citet{uttl05a}, which will be discussed below.

\begin{figure}
\centering
\resizebox{\hsize}{!}{\includegraphics{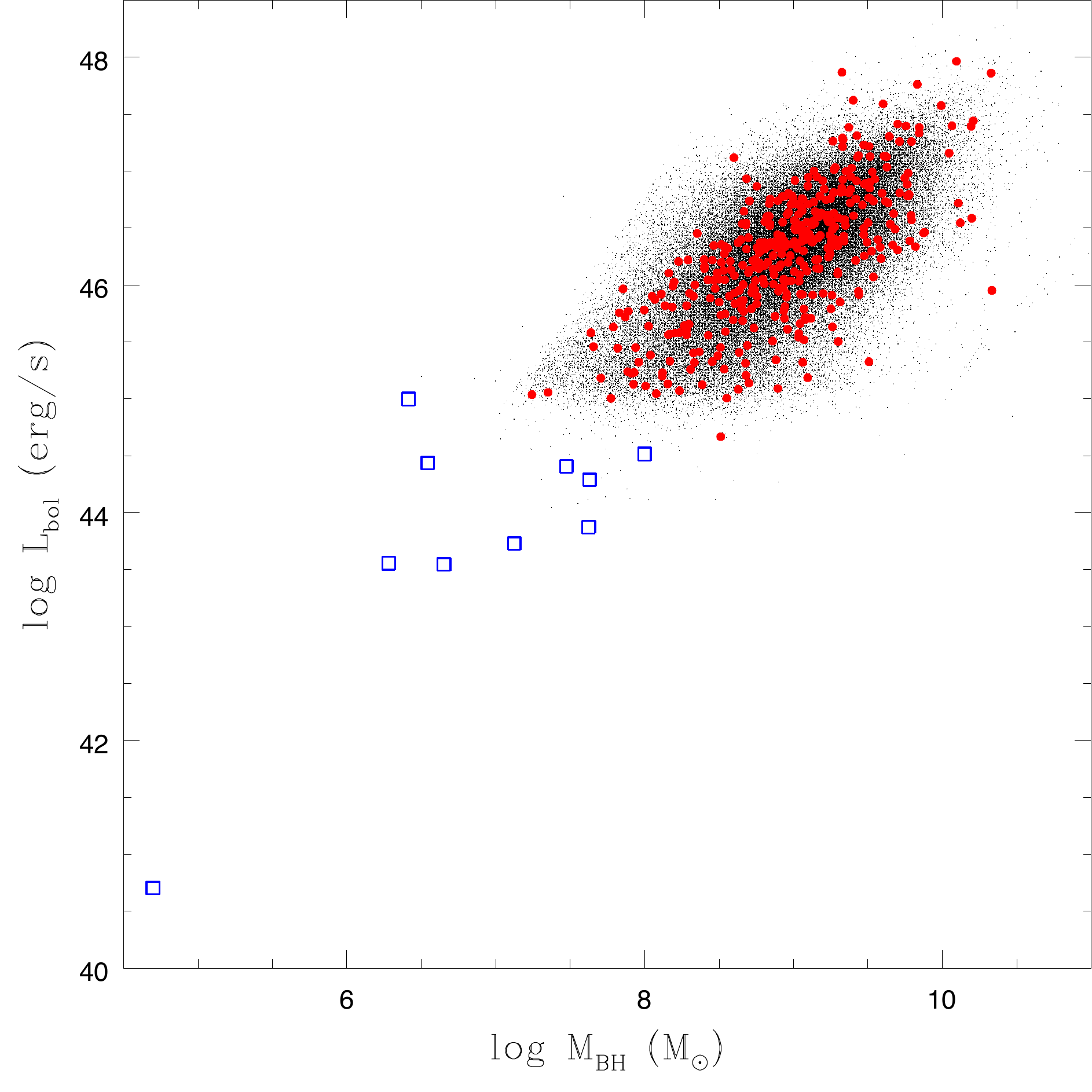}}
\caption{Black hole masses and bolometric luminosities of AGN samples. Black dots: $\sim 100,000$ sources from the \citet{shen11} catalogue. Red filled circles: XMM-Newton sample. Blue empty squares: low-luminosity AGNs from \citet{uttl05a}.}
\end{figure}

We then plot in Figure 9 the structure function for XMM-Newton subsamples binned in intervals of $\log M_{BH}$ and $\log L_{bol}$, with bin width $\Delta\log M_{BH}=\Delta\log L_{bol}=0.5$. The SF is shown for subsamples with at least 30 SF points, in the range of masses $10^{7.5}M_\odot<M_{BH}<10^{10}M_\odot$ and luminosities $10^{45}{\rm\,erg/s}<L_{bol}<10^{47.5}{\rm\,erg/s}$. The total number of SF points is reported in each box, as well as the average SF slope (weighted with the number of points in each bin of $\tau_{rest}$), and the expected value of $\log\tau_{break}$, according to Eq. (6). The SF of the total XMM-Newton sample is also reported for comparison.

\begin{figure}
\centering
\resizebox{\hsize}{!}{\includegraphics{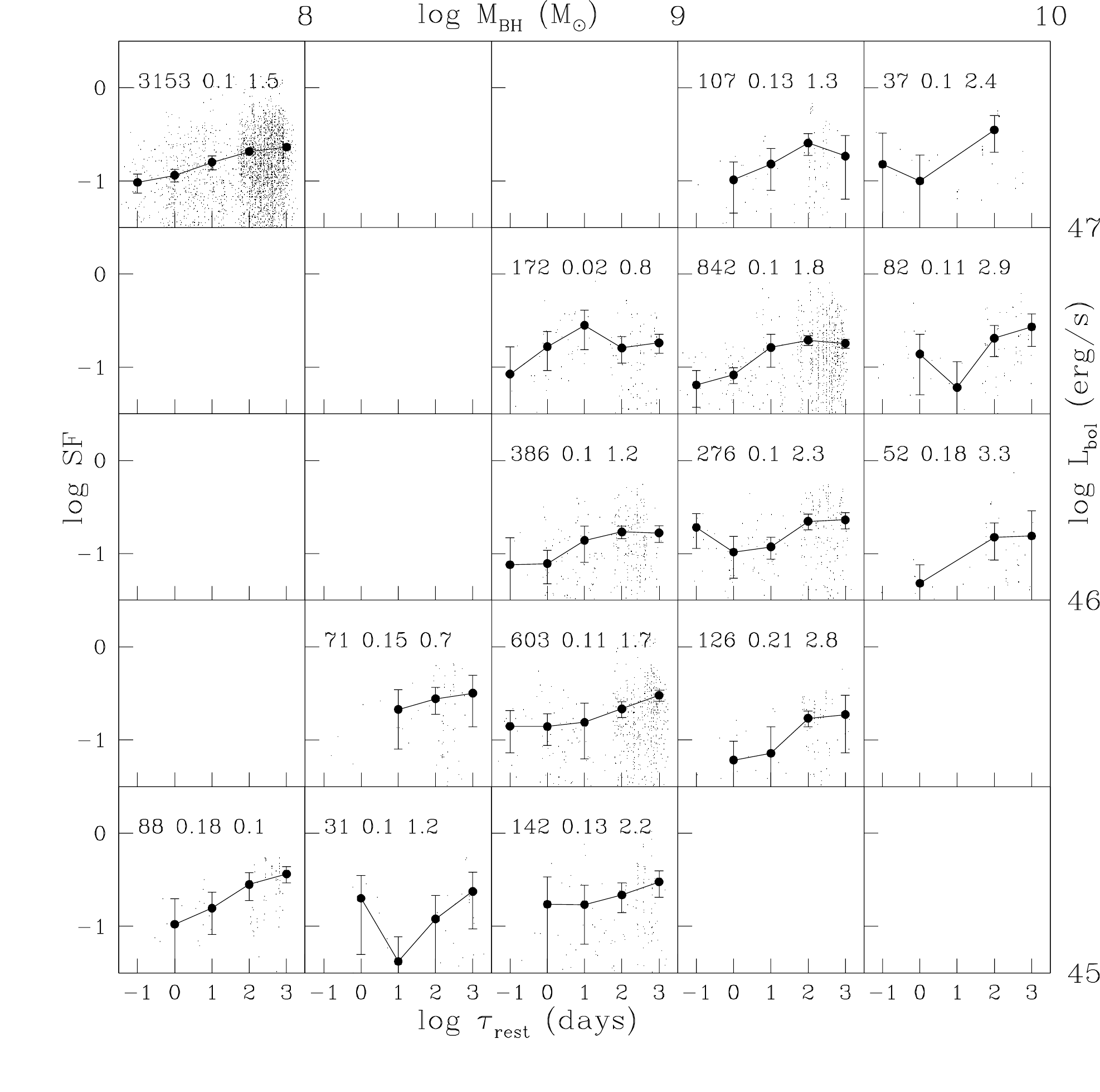}}
\caption{Structure function for XMM-Newton subsamples binned in intervals of $\log M_{BH}$ and $\log L_{bol}$, with bin width $\Delta\log M_{BH}=\Delta\log L_{bol}=0.5$. Values of $M_{BH}$ and $L_{bol}$ are reported in the upper and right axes, respectively. Subsamples with less than 30 SF points are not shown. The SF of the total XMM-Newton sample is reported, for comparison, in the first box in the upper left corner. The values reported in each box are the number of SF points, the average SF slope, and the expected value of $\log\tau_{break}$, according to Eq. (6). Contributions from pairs of individual measurements are also shown (dots).}
\end{figure}

Our results do not support the existence of a break in the SF, expected following Eq. (6). However, we note that the analysis by \citet{mcha06} is based on a few AGNs with quite low luminosities and masses \citep[see][]{uttl05a}, compared to our XMM-Newton sample, and the \citet{shen11} catalogue, see Figure 8.

The absence of a break in our results could be understood if Eq. (6), which appears to hold for AGNs with $M_{BH}\la 10^8M_\odot$ and $L_{bol}\la 10^{45}$ erg/s, would not apply for larger masses and higher luminosities. 

\citet{mcha06} associated the break time scale to a thermal or viscous time scale related to the inner radius of the accretion disk, and identify this with the transition radius $R_{tr}$ predicted by \citet{liu99}, based on evaporation of the inner disk in low Eddington ratio AGNs, describing the transition between an external cool thin disk and an inner, hot, advection-dominated accretion flow (ADAF). This model clearly does not apply to high-luminosity QSOs \citep[see, e.g.][]{nara98}.

\subsection{Dependence on X-ray luminosity and redshift}
Many authors have found inverse dependences of the X-ray variability on the X-ray luminosity $L_X$. Different variability indexes are used, so they must be briefly recalled to make comparisons.

Most authors use the {\it normalised excess variance} \citep[e.g.][]{nand97,vaug03}, defined as
$\sigma^2_{NXS}=(S^2-\sigma_n^2)/{\overline{x}}^2$, where $S^2$ is the total variance of the light curve, $\sigma_n^2$ is the mean square error, and $\overline{x}$ is the mean of N total measurements; or the square root of it, which is also referred to as {\it fractional variability amplitude}, $F_{var}$ \citep[e.g.][]{mark04}.

\citet{gree93} used the {\it normalised variability amplitude}, square root of the power at a specific frequency, normalised to the mean count rate of the related light curve. \citet{lawr93} used the amplitude of the power spectrum at a specific frequency.

As pointed out by \citet{lawr93}, $\sigma^2_{NXS}$ and $F_{var}$ depend on the length of the monitored time interval. Moreover, we notice that they depend on redshift, because the time interval must be properly measured in the rest-frame of the source, as stressed by \citet{gial91} for the optical variability, and by \citet{papa08} for the X-ray case. So the comparison between different results must be taken with some caution. With these limitations in mind, and calling $I_{var}$ a generic variability index (or its square root where appropriate), most of the previous results on the variability dependence on luminosity can be expressed in power-law form, $I_{var}\propto L_X^{-k}$. Values for the exponent $k$ are usually about $\sim 0.3$, for time scales of days, and for samples including Seyfert galaxies and/or low-$z$ QSOs \citep{gree93,lawr93,nand97,mark04}. Similar values are found also for higher redshift QSOs, e.g. by \citet{mann02}, up to $z=2$ ($k=0.27$, still for time scales of days), and by \citet{papa08}, up to $z\sim 3.4$ ($k=0.33$, for time scales of tens of days). Stronger dependences are instead found by \citet{paol04} ($k\sim 0.65$, in the redshift range $0.5<z<1.3$) and by \citet{alma00} ($k=0.75$, for $z<0.5$). For longer time scales (years), a few analyses have been performed, e.g. \citet{mark04} found a weaker dependence, $k\sim 0.13$.

With the analysis of the rest-frame structure function, we can properly compare variability amplitudes at various time lags, and provide an unbiased characterisation of the dependence of variability on luminosity and redshift. In Figure 10 we show the SFs for four luminosity bins between $L_X=10^{43.5}{\rm\,erg/s}$ and $L_X=10^{45.5}{\rm\,erg/s}$: a clear and strong dependence on $L_X$ appears. A change in the slope of the SF is also present (between $\sim$ 0.04 and 0.14), implying that a different dependence on $L_X$ is expected for different time lags. To see this, we re-plot in Figure 11 the SF data vs $L_X$ for two different bins of time lag, centred on 1 day and 100 days, respectively. The least-squares fits, weighted with the number of measurements in each bin, correspond to power-law exponent $k=0.42\pm 0.03$ for the shorter time scale, a slightly stronger dependence, compared to the results by most previous authors. For the longer time scale (100 days), our result is $k=0.21\pm 0.07$, which approximately agrees with the trend found by \citet{mark04}.

\begin{figure}
\centering
\resizebox{\hsize}{!}{\includegraphics{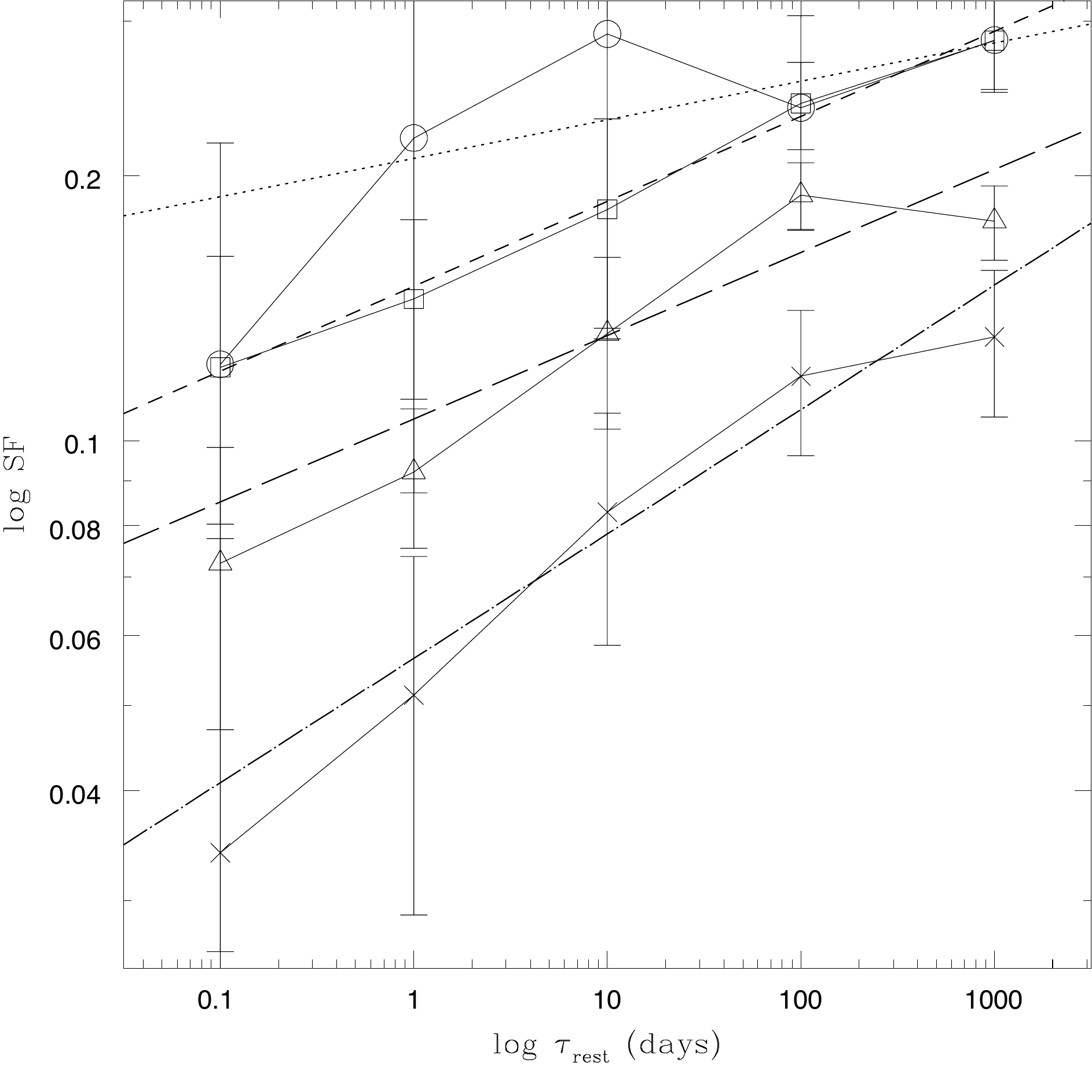}}
\caption{Structure function in bins of X-ray luminosity, represented as points connected by continuous lines. Straight lines with different dash styles: least-squares fits weighted according to the number of points in each bin of time lag. $10^{43.5}{\rm\,erg/s}<L_X<10^{44}{\rm\,erg/s}$: circles, dotted lines; $10^{44}{\rm\,erg/s}<L_X<10^{44.5}{\rm\,erg/s}$: squares, short-dashed lines; $10^{44.5}{\rm\,erg/s}<L_X<10^{45}{\rm\,erg/s}$: triangles, long-dashed lines; $10^{45}{\rm\,erg/s}<L_X<10^{45.5}{\rm\,erg/s}$: crosses, dot-dashed lines.}
\end{figure}

A simple interpretation of the decrease of variability with luminosity ($L$) is the superposition of $N$ randomly flaring subunits. This was already considered in early studies of optical variability \citep[e.g.][]{pica83, aret97}, and, in its simplest version of independent and identical flares, would predict a variability amplitude $\propto N^{-1/2}\propto L^{-1/2}$. In the X-ray domain, several authors have also considered the same argument \citep{gree93,nand97,alma00,mann02}. The observed shallower slope can be understood invoking a correlation among flares \citep[e.g.][]{gree93}, or a dependence of the amplitude of the flares on the luminosity of the source \citep{alma00}. We stress that a simple scaling of the flare amplitude with the luminosity of the source cannot account for the change in the slope of the SF with luminosity, shown in Figure 10, unless some correlation among the flares is also introduced.

Instead of multiple flaring subunits, models based on the variability of a single region have also been considered, e.g. \citet{alma00} explained the dependence of variability on luminosity, invoking a relation between the luminosity and the size of the varying region, which produces a shift of the PSD in the frequency direction, with unchanged slope, under the assumption of self-similar scaling of the variable region. However, a PSD with slope independent on luminosity is inconsistent with our results on the SF (see Fig. 10), implying that a deviation from self-similarity should be considered.

\begin{figure}
\centering
\resizebox{\hsize}{!}{\includegraphics{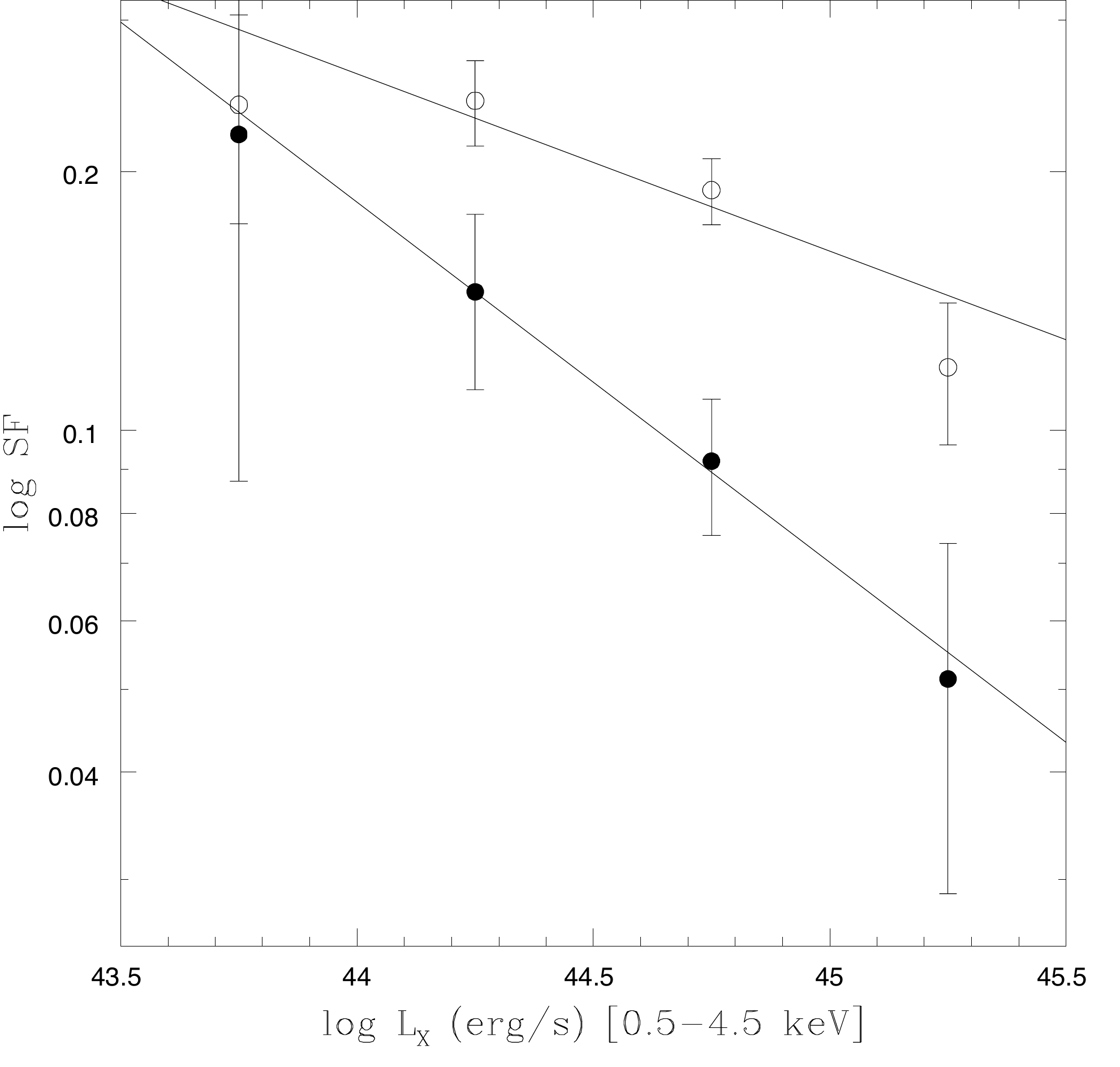}}
\caption{Dependence of the SF on $L_X$. Filled circles: $\tau_{rest}=1$ d; open circles: $\tau_{rest}=100$ d. Lines: weighted least-squares fits, according to the number of points in each bin of $L_X$.}
\end{figure}

The stronger dependences on $L_X$ found by \citet{alma00} and \citet{paol04} are accompanied by the suggestion of a possible increase of the variability with redshift. \citet{alma00} find an opposite dependence on luminosity ($k=-0.3$) for sources at $z>0.5$, which could be caused by an increase with $z$. \citet{paol04} measure a higher variability for sources at $z>1.3$ than for their low-$z$ counterparts of similar luminosity. 
\citet{mann02} also reported  tentative evidence of a stronger variability for sources at $z>2$. Finally, \citet{papa08} compared the variability of high-redshift AGNs in the Lockman Hole region with that of nearby AGNs by \cite{mark04}, finding evidence of an increase with redshift.

Owing to the strong correlation of sources in the $L_X$-$z$ plane (Fig. 1), we limited our analysis of the $z$-dependence to the sources in the luminosity interval $10^{44}{\rm\,erg/s}<L_X<10^{45}{\rm\,erg/s}$, and divided the sample into four equally populated redshift bins, $0<z\leq 1$, $1<z\leq 1.4$, $1.4<z\leq 1.8$, $1.8<z\la 4.5$. The result, displayed in Figure 12, suggests the presence of a weak trend with redshift at intermediate time scales, while at short and long timescales the behaviour appears unclear and possibly non-monotonic.

\begin{figure}
\centering
\resizebox{\hsize}{!}{\includegraphics{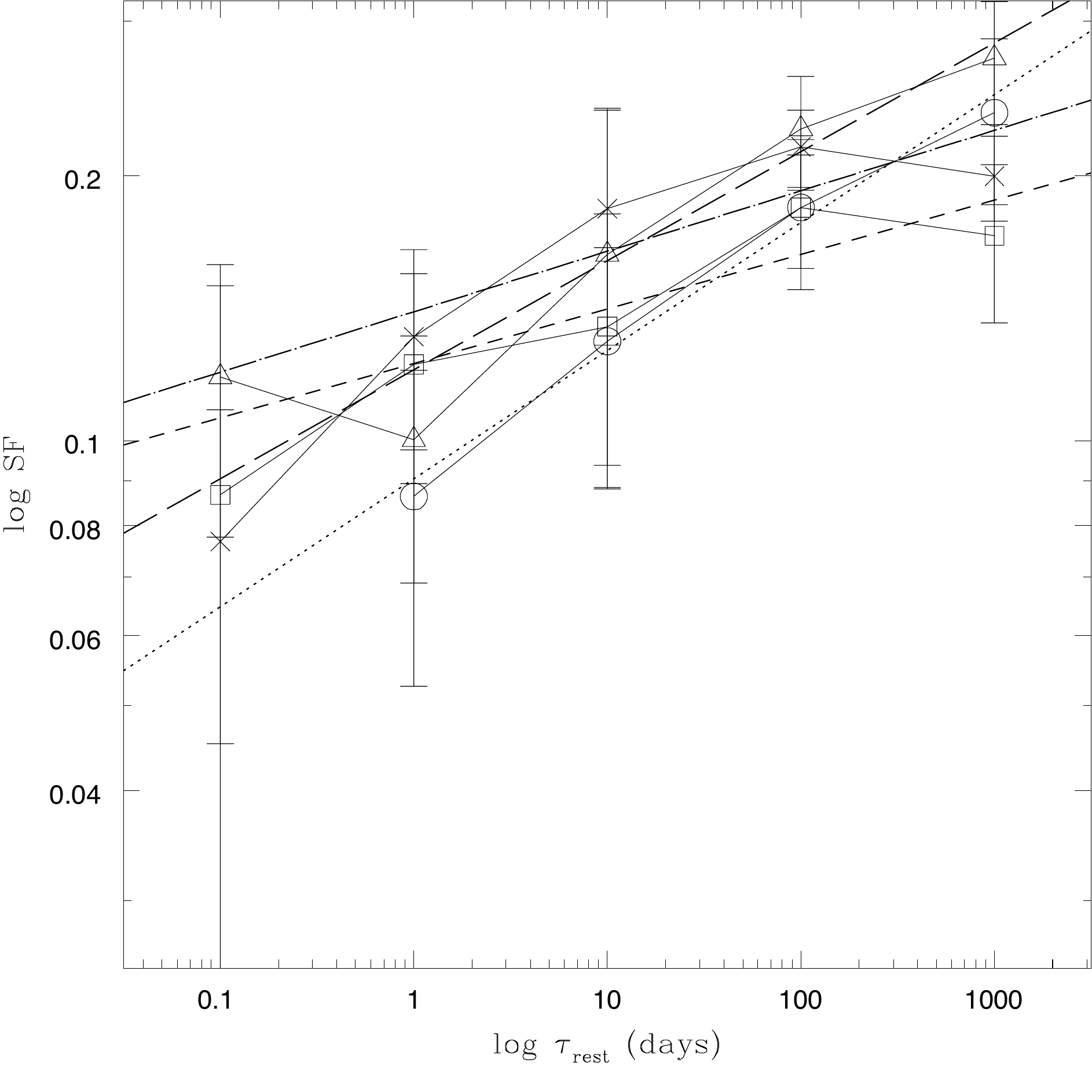}}
\caption{Structure function in bins of redshift for sources in the luminosity interval $10^{44}{\rm\,erg/s}<L_X<10^{45}$ erg/s, represented as points connected by continuous lines. Straight lines with different dash styles: least-squares fits, weighted according to the number of points in each bin of time lag. $0<z\leq 1$: circles, dotted lines; $1<z\leq 1.4$: squares, short-dashed lines; $1.4<z\leq 1.8$: triangles, long-dashed lines; $1.8<z\leq 4.5$: crosses, dot-dashed lines.}
\end{figure}

To investigate this dependence in more detail, we computed partial correlation coefficients of variability with redshift, considering all individual variations that contribute to the SF. While the ordinary correlation coefficient indicates no correlation, $r_{Vz}=-0.06$, with probability $P(>r)=0.001$, the first-order partial correlation coefficient, which takes account of the dependence on $L_X$, is

\begin{equation}
r_{Vz,L}=(r_{Vz}-r_{VL}r_{zL})/\sqrt{(1-r_{VL}^2)(1-r_{zL}^2)}=0.125~,
\end{equation}

\noindent
with probability $P(>r)=10^{-12}$, suggesting the presence of a weak, intrinsic correlation. We also calculated the second-order partial correlation coefficient \citep{kend77}, which compensates for both the dependences on $L_X$ and on the time lag $\tau$, and still strengthens the correlation:

\begin{equation}
r_{Vz,L\tau}=(r_{Vz,L}-r_{V\tau,L}r_{z\tau,L})/\sqrt{(1-r_{V\tau,L}^2)(1-r_{z\tau,L}^2)}=0.127~ .
\end{equation}

\noindent
The probability is in this case $P(>r)=6\cdot 10^{-13}$.

\section{Discussion}
The analysis of X-ray variability of AGNs has previously been performed mainly for individual nearby Seyferts or for small samples of them, and only a few works extend the study to large samples in wide ranges of luminosity and redshift \citep[e.g.][]{alma00,mann02,paol04}. Our study presents the first ensemble analysis based on the structure function. This is to be preferred for statistic studies compared with a PSD analysis, because SF operates in the time domain, is less dependent on irregular sampling, and allows for an analysis even with very few epochs. The SF is also preferable compared with the analysis of fractional variability and excess variance, because these parameters are biased by the duration of the monitoring time interval in the rest-frame, and thus on cosmological time dilation.

Our variability analysis, based on two different serendipitously selected samples extracted from the catalogues of XMM-Newton \citep{wats09} and Swift \citep{pucc11}, gives statistically consistent results in the two cases, with the SF described by a power law of the time lag, with exponent $b=0.10\pm 0.01$ (XMM-Newton) or $b=0.07\pm 0.04$ (Swift). This would correspond to a PSD with power law exponent $\alpha\approx 1.2$ for the case of a single-power-law PSD, which is within the range of exponents found for nearby Seyferts \citep{lawr93}.

While the PSD of local low-luminosity AGNs often shows one or two breaks, we do not find evidence of breaks in the SF, even dividing the analysis in bins of $M_{BH}$ and $L_{bol}$. However, while a break at a time lag roughly proportional to the black hole mass is expected for local AGNs, our results do not support this expectation for more luminous AGNs and QSOs. This suggests that the relation found by \citet{mcha06}, reported in Eq. (6), cannot be extrapolated to high bolometric luminosities and large black hole masses, possibly because the transition between an external cool thin disk and an inner ADAF \citep{liu99} does not apply in the high-Eddington ratio regime.

We confirm a strong anti-correlation of the variability with X-ray luminosity, as $L_X^{-0.42}$ and as $L_X^{-0.21}$ for time lags $\sim 1$ day and $\sim 100$ days, respectively. This approximately agrees with most previous authors \citep{gree93,lawr93,nand97,mark04,papa08}.

The behaviour of the slope and amplitude of the SF as a function of the luminosity implies that (i) for a model of multiple flaring subunits, they cannot be uncorrelated, (ii) for a model with a single varying region self-similar scaling with luminosity cannot hold.

We find evidence in support of a weak, intrinsic, increase of the average X-ray variability with redshift. The dependence, however, appears tangled with that on the time lag. This suggests that different  processes could dominate the variability at short and long time scales, and that their relative importance changes with the redshift.

\begin{acknowledgements}
We are grateful to Paolo Giommi, Maurizio Paolillo, Matteo Perri, and Simonetta Puccetti for useful discussions. S.T. acknowledges financial support through Grant ASI I/088/06/0. Part of this work is based on archival data, software or on-line services provided by the ASI Science Data Center (ASDC). This research made use of the XMM-Newton Serendipitous Source Catalogue, which is a collaborative project involving the whole Science Survey Center Consortium. Funding for the SDSS and SDSS-II was provided by the Alfred P. Sloan Foundation, the Participating Institutions, the National Science Foundation, the U.S. Department of Energy, the National Aeronautics and Space Administration, the Japanese Monbukagakusho, the Max Planck Society, and the Higher Education Funding Council for England. The SDSS was managed by the Astrophysical Research Consortium for the Participating Institutions.
\end{acknowledgements}

\bibliographystyle{aa}
\bibliography{xvar2.bib}{}

\onltab{1}{
\begin{longtable}{rllrccrr}
\caption{XMM-Newton sample.}\\
\hline\hline
$N_{sou}$ &name &$z$ &$N_{epo}$ &$\log f_X$ &$\log L_X$ &$\log\tau_{min}$ &$\log\tau_{max}$ \\
(1) &(2)  &(3)  &(4) &(5)  &(6)  &(7)  &(8) \\
\hline
\endfirsthead
\caption{continued.}\\
\hline\hline
$N_{sou}$ &name &$z$ &$N_{epo}$ &$\log f_X$ &$\log L_X$ &$\log\tau_{min}$ &$\log\tau_{max}$ \\
(1) &(2)  &(3)  &(4) &(5)  &(6)  &(7)  &(8) \\
\hline
\endhead
\hline
\endfoot
   1 & 2XMMi J001716.8-010725 &  1.163 &  2 &  -12.72 &  45.06 &  1.10 & 1.10 \\
   2 & 2XMMi J001808.7-005709 &  1.335 &  2 &  -13.39 &  44.53 &  1.06 & 1.06 \\
   3 & 2XMM  J020011.5-093125 & 0.3604 &  2 &  -12.49 &  44.11 &  2.15 & 2.15 \\
   4 & 2XMM  J020118.6-091936 & 0.6607 &  2 &  -12.73 &  44.48 &  2.06 & 2.06 \\
   5 & 2XMM  J024040.8-081309 &   1.85 &  2 &  -13.92 &  44.32 &  2.75 & 2.75 \\
   6 & 2XMM  J024055.8-081952 &  1.801 &  2 &  -13.84 &  44.37 &  2.76 & 2.76 \\
   7 & 2XMM  J024105.8-081153 & 0.9785 &  2 &  -13.66 &  43.95 &  2.91 & 2.91 \\
   8 & 2XMM  J024125.9-080936 &  3.072 &  2 &  -14.13 &  44.60 &  2.60 & 2.60 \\
   9 & 2XMM  J024149.9-000433 &   1.26 &  2 &  -13.46 &  44.40 & -0.43 & -0.43 \\
  10 & 2XMM  J024157.1+000703 &  1.563 &  2 &  -13.46 &  44.62 & -0.48 & -0.48 \\
  11 & 2XMM  J024200.8+000021 &  1.104 &  2 &  -12.85 &  44.88 & -0.39 & -0.39 \\
  12 & 2XMM  J024204.7+000814 & 0.3822 &  2 &  -13.25 &  43.41 & -0.21 & -0.21 \\
  13 & 2XMM  J024207.2+000038 & 0.3842 &  2 &  -13.06 &  43.61 & -0.21 & -0.21 \\
  14 & 2XMM  J024215.0-000209 &   1.01 &  2 &  -13.62 &  44.02 & -0.37 & -0.37 \\
  15 & 2XMM  J024227.3+000846 & 0.6501 &  2 &  -13.45 &  43.74 & -0.29 & -0.29 \\
  16 & 2XMM  J024250.8-000030 &  2.177 &  2 &  -14.08 &  44.32 & -0.57 & -0.57 \\
  17 & 2XMM  J024251.0+001010 &  1.888 &  2 &  -13.27 &  45.00 & -0.53 & -0.53 \\
  18 & 2XMM  J024304.6+000005 &  1.995 &  2 &  -13.51 &  44.80 & -0.55 & -0.55 \\
  19 & 2XMM  J024308.1-000126 & 0.6787 &  2 &  -13.67 &  43.57 & -0.30 & -0.30 \\
  20 & 2XMM  J030707.3-000424 & 0.6641 &  2 &  -13.20 &  44.02 &  2.50 & 2.50 \\
  21 & 2XMM  J032108.4+413221 &  2.467 &  2 &  -13.39 &  45.13 &  2.32 & 2.32 \\
  22 & 2XMM  J033627.4+004653 &  1.746 &  3 &  -13.96 &  44.22 &  2.15 & 2.53 \\
  23 & 2XMM  J033639.5+002535 &   1.68 &  9 &  -13.31 &  44.84 & -0.22 & 2.79 \\
  24 & 2XMM  J033654.2+004015 &  2.625 &  6 &  -13.42 &  45.16 & -0.35 & 2.66 \\
  25 & 2XMM  J033701.1+004312 &  2.006 &  3 &  -14.01 &  44.31 &  2.11 & 2.49 \\
  26 & 2XMM  J033709.1+004614 &  2.506 &  4 &  -14.17 &  44.37 & -0.34 & 2.42 \\
  27 & 2XMM  J033711.5+004344 &  1.918 &  3 &  -14.18 &  44.10 &  1.79 & 2.40 \\
  28 & 2XMM  J033715.6+004206 &  2.354 &  6 &  -14.00 &  44.48 & -0.32 & 2.44 \\
  29 & 2XMM  J033716.5+003124 &  2.437 &  3 &  -14.54 &  43.97 &  1.71 & 2.03 \\
  30 & 2XMM  J033718.8+003303 & 0.4371 &  7 &  -13.80 &  42.99 &  0.05 & 2.81 \\
  31 & 2XMM  J033746.7+003510 &    1.4 &  2 &  -13.78 &  44.19 &  0.81 & 0.81 \\
  32 & 2XMM  J033754.1+002934 &  2.004 &  2 &  -13.43 &  44.89 &  2.39 & 2.39 \\
  33 & 2XMM  J033801.9+002719 &  1.583 &  2 &  -13.68 &  44.41 &  2.45 & 2.45 \\
  34 & 2XMM  J073405.2+320315 &  2.082 &  2 &  -14.13 &  44.23 &  1.71 & 1.71 \\
  35 & 2XMMi J073654.0+302657 & 0.7238 &  2 &  -13.13 &  44.17 &  1.99 & 1.99 \\
  36 & 2XMMi J073708.1+303914 &  1.403 &  2 &  -13.36 &  44.61 &  1.85 & 1.85 \\
  37 & 2XMMi J073712.4+303637 & 0.9177 &  2 &  -12.73 &  44.81 &  1.95 & 1.95 \\
  38 & 2XMM  J074222.3+494147 & 0.9274 &  6 &  -12.97 &  44.58 &  1.22 & 3.16 \\
  39 & 2XMM  J080633.2+153810 & 0.9994 &  2 &  -13.34 &  44.29 &  2.57 & 2.57 \\
  40 & 2XMM  J083102.9+523534 &  4.444 &  3 &  -14.26 &  44.81 &  0.47 & 2.57 \\
  41 & 2XMM  J083740.2+245423 &  1.125 &  2 &  -12.38 &  45.37 &  1.93 & 1.93 \\
  42 & 2XMMi J083906.7+575417 &  1.534 &  2 &  -12.23 &  45.83 &  1.94 & 1.94 \\
  43 & 2XMMi J083924.8+575231 &  0.187 &  2 &  -12.52 &  43.45 &  2.27 & 2.27 \\
  44 & 2XMM  J084659.3+344825 &  1.582 &  2 &  -13.23 &  44.86 &  2.12 & 2.12 \\
  45 & 2XMM  J084710.0+345442 &  2.303 &  2 &  -13.63 &  44.82 &  2.01 & 2.01 \\
  46 & 2XMM  J084905.0+445714 &  1.259 &  3 &  -13.44 &  44.41 & -0.56 & 0.46 \\
  47 & 2XMM  J084943.6+450023 &  1.593 &  3 &  -13.53 &  44.56 & -0.62 & 0.40 \\
  48 & 2XMM  J085346.1+200957 &  1.093 &  2 &  -13.18 &  44.53 &  2.26 & 2.26 \\
  49 & 2XMMi J090429.5+340544 &  1.297 &  2 &  -13.67 &  44.22 &  2.19 & 2.19 \\
  50 & 2XMMi J090505.5+341352 &  1.024 &  2 &  -12.78 &  44.87 &  2.25 & 2.25 \\
  51 & 2XMMi J090516.6+340921 &  1.872 &  2 &  -13.35 &  44.90 &  2.09 & 2.09 \\
  52 & 2XMMi J090525.2+341500 &  1.588 &  2 &  -13.58 &  44.51 &  2.14 & 2.14 \\
  53 & 2XMM  J091301.0+525929 &  1.377 &  2 &  -12.09 &  45.86 &  1.11 & 1.11 \\
  54 & 2XMM  J091302.8+530322 & 0.6307 &  2 &  -13.98 &  43.18 &  1.28 & 1.28 \\
  55 & 2XMM  J091440.3+530038 &   1.43 &  2 &  -13.60 &  44.38 &  1.11 & 1.11 \\
  56 & 2XMM  J091843.6+211819 & 0.8309 &  2 &  -13.45 &  43.99 &  2.71 & 2.71 \\
  57 & 2XMM  J091848.6+211717 & 0.1493 &  2 &  -12.47 &  43.29 &  2.91 & 2.91 \\
  58 & 2XMM  J091852.9+211518 &  1.035 &  2 &  -14.33 &  43.33 &  2.66 & 2.66 \\
  59 & 2XMM  J091907.5+212553 &   1.39 &  2 &  -12.90 &  45.06 &  2.59 & 2.59 \\
  60 & 2XMM  J091908.7+212153 &  1.514 &  2 &  -13.32 &  44.72 &  2.57 & 2.57 \\
  61 & 2XMM  J091914.2+303018 &  1.388 &  2 &  -14.11 &  43.84 &  2.78 & 2.78 \\
  62 & 2XMM  J092039.7+301701 &   1.18 &  2 &  -14.00 &  43.79 & -0.98 & -0.98 \\
  63 & 2XMM  J092104.3+302031 &   3.35 &  3 &  -13.80 &  45.01 & -1.28 & 2.52 \\
  64 & 2XMM  J093359.2+551550 &  1.863 &  2 &  -13.26 &  44.99 &  0.33 & 0.33 \\
  65 & 2XMM  J093551.5+551117 &   1.79 &  2 &  -13.63 &  44.58 &  0.34 & 0.34 \\
  66 & 2XMM  J093555.4+551238 &    1.8 &  2 &  -13.33 &  44.89 &  0.34 & 0.34 \\
  67 & 2XMM  J094404.3+480647 & 0.3919 &  2 &  -12.74 &  43.94 &  1.32 & 1.32 \\
  68 & 2XMM  J095251.5+013848 & 0.4997 &  2 &  -13.39 &  43.53 &  2.05 & 2.05 \\
  69 & 2XMM  J095344.9+014251 &  1.657 &  2 &  -13.22 &  44.91 &  1.80 & 1.80 \\
  70 & 2XMM  J095636.3+690028 &  1.975 &  3 &  -13.47 &  44.84 &  0.30 & 2.48 \\
  71 & 2XMM  J095658.6+693852 &  2.035 &  2 &  -13.38 &  44.96 &  2.55 & 2.55 \\
  72 & 2XMM  J095701.3+685500 &  1.297 &  3 &  -13.07 &  44.82 &  0.42 & 2.59 \\
  73 & 2XMMi J095750.0+013352 &  2.011 &  2 &  -13.88 &  44.45 &  1.00 & 1.00 \\
  74 & 2XMM  J095754.7+023831 &    1.6 &  2 &  -14.14 &  43.96 &  1.14 & 1.14 \\
  75 & 2XMMi J095759.4+020435 &  2.034 &  3 &  -13.81 &  44.53 &  0.96 & 2.09 \\
  76 & 2XMM  J095810.9+014004 &  2.101 &  3 &  -13.64 &  44.73 &  0.99 & 2.10 \\
  77 & 2XMM  J095815.5+014922 &  1.509 &  6 &  -12.92 &  45.12 & -1.01 & 2.20 \\
  78 & 2XMM  J095819.8+022903 & 0.3454 &  3 &  -13.08 &  43.48 &  1.43 & 2.62 \\
  79 & 2XMMi J095820.5+020303 &  1.355 &  3 &  -14.26 &  43.67 &  1.11 & 2.20 \\
  80 & 2XMM  J095822.2+014524 &   1.96 &  5 &  -13.57 &  44.73 & -1.08 & 2.13 \\
  81 & 2XMM  J095834.0+024427 &  1.887 &  2 &  -13.46 &  44.80 &  2.09 & 2.09 \\
  82 & 2XMMi J095834.7+014502 &  1.889 &  5 &  -13.99 &  44.27 & -1.07 & 2.10 \\
  83 & 2XMM  J095844.9+014309 &  1.337 &  5 &  -13.93 &  43.99 &  0.38 & 2.49 \\
  84 & 2XMM  J095847.7+690533 &  1.288 &  3 &  -12.82 &  45.06 &  0.42 & 2.59 \\
  85 & 2XMM  J095848.8+023441 &  1.549 &  6 &  -13.54 &  44.53 & -0.18 & 2.34 \\
  86 & 2XMM  J095852.1+025156 &  1.407 &  2 &  -13.31 &  44.66 &  2.17 & 2.17 \\
  87 & 2XMM  J095857.3+021314 &  1.024 &  5 &  -12.75 &  44.90 &  0.44 & 2.79 \\
  88 & 2XMM  J095858.6+020139 &  2.456 & 10 &  -13.41 &  45.11 & -0.95 & 2.56 \\
  89 & 2XMM  J095902.7+021906 & 0.3454 &  9 &  -12.61 &  43.95 & -0.72 & 2.97 \\
  90 & 2XMM  J095908.3+024309 &  1.318 &  2 &  -12.72 &  45.18 &  2.18 & 2.18 \\
  91 & 2XMM  J095918.7+020951 &  1.156 &  7 &  -12.85 &  44.93 & -0.09 & 2.76 \\
  92 & 2XMM  J095924.4+015954 &  1.235 &  8 &  -13.09 &  44.75 & -0.10 & 2.61 \\
  93 & 2XMM  J095935.6+024838 &  1.973 &  2 &  -14.23 &  44.08 &  2.08 & 2.08 \\
  94 & 2XMM  J095946.0+024743 &  1.066 &  3 &  -13.12 &  44.57 &  1.90 & 2.41 \\
  95 & 2XMM  J095949.4+020141 &  1.753 &  8 &  -13.62 &  44.57 & -0.84 & 2.59 \\
  96 & 2XMM  J095958.0+014327 &  1.627 &  5 &  -13.93 &  44.19 & -0.02 & 2.53 \\
  97 & 2XMM  J100001.3+024845 & 0.7661 &  3 &  -13.20 &  44.16 &  1.97 & 2.47 \\
  98 & 2XMM  J100008.0+013307 &  1.172 &  4 &  -13.80 &  43.99 &  1.91 & 2.62 \\
  99 & 2XMM  J100012.9+023522 & 0.6984 &  6 &  -13.11 &  44.16 &  0.51 & 2.72 \\
 100 & 2XMM  J100014.1+020054 &  2.498 &  7 &  -13.75 &  44.78 & -0.95 & 2.49 \\
 101 & 2XMM  J100024.3+015053 &  1.664 &  5 &  -13.85 &  44.29 &  0.35 & 2.61 \\
 102 & 2XMM  J100024.6+023148 &  1.321 &  5 &  -13.40 &  44.51 &  0.38 & 2.58 \\
 103 & 2XMM  J100025.2+015852 & 0.3726 &  7 &  -12.53 &  44.11 & -0.54 & 2.90 \\
 104 & 2XMM  J100043.1+020637 &   0.36 & 10 &  -13.23 &  43.37 & -0.54 & 2.90 \\
 105 & 2XMM  J100055.4+023442 &  1.404 &  8 &  -13.53 &  44.43 & -0.97 & 2.58 \\
 106 & 2XMM  J100058.8+015359 &  1.557 & 10 &  -13.55 &  44.53 & -1.11 & 2.63 \\
 107 & 2XMM  J100104.2+553522 &  1.535 &  2 &  -13.31 &  44.74 &  2.56 & 2.56 \\
 108 & 2XMM  J100114.3+022356 &  1.796 &  6 &  -13.56 &  44.65 & -1.04 & 2.41 \\
 109 & 2XMM  J100116.7+014053 &  2.054 &  4 &  -13.75 &  44.59 & -0.12 & 2.25 \\
 110 & 2XMM  J100120.2+023341 &  1.834 &  5 &  -13.81 &  44.43 & -1.04 & 2.51 \\
 111 & 2XMM  J100120.7+555351 &  1.413 &  2 &  -11.98 &  45.99 &  2.58 & 2.58 \\
 112 & 2XMM  J100130.3+014304 &   1.57 &  4 &  -14.04 &  44.04 & -0.83 & 2.15 \\
 113 & 2XMM  J100132.2+013419 &  1.355 &  3 &  -13.78 &  44.15 & -0.79 & 2.19 \\
 114 & 2XMM  J100136.4+025304 &  2.116 &  3 &  -13.54 &  44.84 &  1.79 & 2.37 \\
 115 & 2XMM  J100145.2+022456 &  2.032 &  3 &  -14.00 &  44.33 &  0.90 & 2.09 \\
 116 & 2XMM  J100156.3+555440 &  1.152 &  2 &  -13.33 &  44.44 &  2.63 & 2.63 \\
 117 & 2XMM  J100201.5+020330 &  2.023 &  6 &  -14.20 &  44.12 & -0.42 & 2.27 \\
 118 & 2XMM  J100202.7+022434 & 0.9877 &  8 &  -13.57 &  44.05 &  0.56 & 2.66 \\
 119 & 2XMM  J100205.2+554258 &  1.151 &  2 &  -13.59 &  44.18 &  2.63 & 2.63 \\
 120 & 2XMM  J100210.6+023026 &  1.161 &  6 &  -13.41 &  44.36 & -0.74 & 2.63 \\
 121 & 2XMM  J100219.5+015537 &   1.51 &  7 &  -13.60 &  44.44 & -0.82 & 2.34 \\
 122 & 2XMM  J100226.3+021923 &  1.292 &  5 &  -13.38 &  44.50 &  0.50 & 2.38 \\
 123 & 2XMM  J100232.1+023537 & 0.6576 &  4 &  -13.06 &  44.14 & -0.62 & 2.52 \\
 124 & 2XMM  J100234.3+015011 &  1.504 &  4 &  -13.37 &  44.67 & -0.81 & 2.16 \\
 125 & 2XMM  J100236.6+015949 &  1.516 &  2 &  -14.07 &  43.97 &  1.84 & 1.84 \\
 126 & 2XMM  J100238.2+013747 &  2.506 &  2 &  -13.63 &  44.91 &  2.02 & 2.02 \\
 127 & 2XMM  J100243.5+324812 & 0.7116 &  2 &  -13.17 &  44.12 &  2.52 & 2.52 \\
 128 & 2XMM  J100248.9+325130 &  1.537 &  2 &  -13.13 &  44.93 &  2.35 & 2.35 \\
 129 & 2XMMi J100251.6+022905 &  2.006 &  2 &  -13.81 &  44.51 &  2.26 & 2.26 \\
 130 & 2XMM  J100254.4+324039 & 0.8288 &  2 &  -12.05 &  45.39 &  2.49 & 2.49 \\
 131 & 2XMM  J100302.9+015208 &    1.8 &  4 &  -13.36 &  44.85 & -0.86 & 2.12 \\
 132 & 2XMM  J100309.2+022037 &  1.964 &  3 &  -13.91 &  44.39 &  0.39 & 2.12 \\
 133 & 2XMM  J100309.4+554134 & 0.6736 &  2 &  -12.95 &  44.28 &  2.74 & 2.74 \\
 134 & 2XMM  J100324.5+021830 & 0.5184 &  2 &  -12.79 &  44.17 &  2.40 & 2.40 \\
 135 & 2XMM  J100926.6+533424 &   1.73 &  2 &  -13.02 &  45.16 &  2.94 & 2.94 \\
 136 & 2XMMi J102129.1+215609 &  1.465 &  2 &  -13.35 &  44.67 &  1.82 & 1.82 \\
 137 & 2XMMi J102134.2+215437 &  1.536 &  2 &  -13.49 &  44.57 &  1.81 & 1.81 \\
 138 & 2XMMi J102223.7+383424 &  1.357 &  2 &  -13.62 &  44.31 &  0.82 & 0.82 \\
 139 & 2XMMi J102224.0+215832 &  1.165 &  2 &  -13.64 &  44.14 &  1.88 & 1.88 \\
 140 & 2XMMi J102255.3+383007 &  0.658 &  2 &  -13.57 &  43.64 &  0.98 & 0.98 \\
 141 & 2XMM  J102310.0+194248 &  1.594 &  2 &  -13.64 &  44.45 &  2.93 & 2.93 \\
 142 & 2XMM  J102313.2+195651 &  1.086 &  2 &  -13.13 &  44.58 &  3.02 & 3.02 \\
 143 & 2XMM  J102318.6+194835 &  1.761 &  2 &  -13.77 &  44.42 &  2.90 & 2.90 \\
 144 & 2XMM  J102423.7+195250 &  1.635 &  2 &  -13.83 &  44.29 &  2.92 & 2.92 \\
 145 & 2XMM  J103216.0+505119 & 0.1731 &  4 &  -12.44 &  43.46 &  0.71 & 3.03 \\
 146 & 2XMM  J103227.9+573822 &  1.969 &  2 &  -13.23 &  45.07 &  1.15 & 1.15 \\
 147 & 2XMM  J103518.5+392934 & 0.8774 &  2 &  -13.61 &  43.89 &  3.00 & 3.00 \\
 148 & 2XMM  J104401.1+212804 &  1.494 &  3 &  -13.17 &  44.86 &  0.98 & 1.95 \\
 149 & 2XMM  J104414.5+213203 &   1.17 &  3 &  -13.65 &  44.13 &  1.04 & 2.01 \\
 150 & 2XMM  J104440.0+212643 &  1.504 &  3 &  -13.34 &  44.70 &  0.97 & 1.95 \\
 151 & 2XMM  J104522.1+212614 & 0.8908 &  3 &  -13.07 &  44.44 &  1.10 & 2.07 \\
 152 & 2XMM  J105039.5+572336 &  1.445 &  3 &  -13.35 &  44.64 & -0.09 & 0.39 \\
 153 & 2XMM  J105050.0+573819 &  1.285 &  3 &  -13.33 &  44.55 & -0.06 & 0.42 \\
 154 & 2XMM  J105201.3+441417 &   1.79 &  2 &  -13.72 &  44.48 &  1.02 & 1.02 \\
 155 & 2XMM  J105204.5+440152 &  1.524 &  2 &  -13.47 &  44.58 &  1.06 & 1.06 \\
 156 & 2XMM  J105221.0+440439 & 0.9677 &  2 &  -13.16 &  44.43 &  1.17 & 1.17 \\
 157 & 2XMM  J105224.9+441505 & 0.4435 &  2 &  -12.48 &  44.33 &  1.31 & 1.31 \\
 158 & 2XMM  J105239.6+572431 &  1.112 & 14 &  -12.90 &  44.83 & -0.05 & 2.65 \\
 159 & 2XMM  J105316.7+573550 &  1.205 & 13 &  -12.78 &  45.03 & -0.09 & 2.64 \\
 160 & 2XMM  J105404.1+574019 &  1.102 &  3 &  -13.45 &  44.27 & -0.07 & 0.26 \\
 161 & 2XMM  J105422.5+572031 &  2.972 &  3 &  -13.94 &  44.76 & -0.35 & -0.02 \\
 162 & 2XMMi J105540.0+065552 &  0.596 &  5 &  -13.46 &  43.64 &  0.08 & 2.01 \\
 163 & 2XMMi J105549.4+065542 & 0.9024 &  5 &  -12.77 &  44.75 &  0.01 & 1.93 \\
 164 & 2XMMi J105603.7+070235 &  2.303 &  5 &  -13.38 &  45.07 & -0.23 & 1.70 \\
 165 & 2XMMi J105622.1+071250 &  1.656 &  5 &  -13.09 &  45.05 & -0.14 & 1.79 \\
 166 & 2XMM  J110253.4+360425 &  1.795 &  2 &  -13.90 &  44.31 &  2.59 & 2.59 \\
 167 & 2XMM  J110309.2+380914 &  1.721 &  2 &  -13.44 &  44.73 &  1.02 & 1.02 \\
 168 & 2XMM  J110320.1+380931 &  1.752 & 12 &  -13.76 &  44.43 & -0.87 & 2.83 \\
 169 & 2XMM  J110334.7+355108 &  1.199 &  2 &  -13.75 &  44.06 &  2.70 & 2.70 \\
 170 & 2XMM  J110400.3+380231 &  1.621 & 21 &  -13.74 &  44.37 & -1.31 & 3.05 \\
 171 & 2XMM  J110449.0+381812 &  1.943 & 25 &  -13.54 &  44.75 & -1.36 & 3.05 \\
 172 & 2XMM  J110458.2+250422 &  3.574 &  2 &  -13.48 &  45.39 &  2.20 & 2.20 \\
 173 & 2XMM  J110547.1+380948 &  1.153 &  6 &  -13.12 &  44.65 & -1.22 & 2.93 \\
 174 & 2XMM  J110550.6+251747 & 0.6247 &  2 &  -12.53 &  44.62 &  2.65 & 2.65 \\
 175 & 2XMM  J110602.6+251227 &  1.683 &  2 &  -13.48 &  44.66 &  2.43 & 2.43 \\
 176 & 2XMMi J111233.4+060619 &  3.278 &  2 &  -14.09 &  44.70 & -0.36 & -0.36 \\
 177 & 2XMMi J111303.6+061620 & 0.8491 &  2 &  -13.16 &  44.30 &  0.00 & 0.00 \\
 178 & 2XMM  J111506.0+424949 & 0.3005 &  2 &  -13.16 &  43.27 &  2.83 & 2.83 \\
 179 & 2XMM  J111747.3+075400 &  1.961 &  3 &  -13.88 &  44.42 &  0.72 & 2.50 \\
 180 & 2XMM  J111816.9+074558 &  1.735 &  3 &  -12.42 &  45.76 &  0.76 & 2.54 \\
 181 & 2XMM  J111840.5+075323 &  1.461 &  3 &  -13.37 &  44.63 &  0.80 & 2.58 \\
 182 & 2XMM  J111842.3+212014 &  1.924 &  2 &  -13.61 &  44.67 &  2.58 & 2.58 \\
 183 & 2XMM  J111853.4+074946 &  2.042 &  3 &  -13.66 &  44.68 &  0.71 & 2.49 \\
 184 & 2XMM  J111902.0+213315 &  1.933 &  2 &  -13.39 &  44.90 &  2.58 & 2.58 \\
 185 & 2XMM  J111928.3+130250 &  2.394 &  2 &  -12.81 &  45.68 &  2.03 & 2.03 \\
 186 & 2XMM  J113205.1+530726 &   1.84 &  2 &  -13.74 &  44.50 &  1.82 & 1.82 \\
 187 & 2XMM  J113224.0+525157 &  0.837 &  2 &  -13.38 &  44.07 &  2.00 & 2.00 \\
 188 & 2XMM  J114405.6+195734 & 0.9541 &  2 &  -13.06 &  44.52 &  1.96 & 1.96 \\
 189 & 2XMM  J115606.7+233106 &  1.593 &  3 &  -12.90 &  45.19 &  0.79 & 2.99 \\
 190 & 2XMM  J115726.2+434954 &  1.597 &  2 &  -13.23 &  44.87 & -0.16 & -0.16 \\
 191 & 2XMM  J115838.5+435505 &  1.208 &  2 &  -13.63 &  44.18 & -0.09 & -0.09 \\
 192 & 2XMM  J115851.0+435048 & 0.2871 &  2 &  -12.92 &  43.46 &  0.14 & 0.14 \\
 193 & 2XMM  J115906.3+434643 &  1.462 &  2 &  -13.50 &  44.51 & -0.14 & -0.14 \\
 194 & 2XMM  J115911.3+440819 &  1.438 &  2 &  -13.50 &  44.50 & -0.13 & -0.13 \\
 195 & 2XMM  J120405.8+201345 & 0.5985 &  3 &  -12.83 &  44.27 &  1.10 & 2.05 \\
 196 & 2XMM  J120414.4+351759 &  2.359 &  2 &  -13.86 &  44.62 &  1.05 & 1.05 \\
 197 & 2XMM  J120432.7+202434 &   2.09 &  3 &  -13.34 &  45.02 &  0.81 & 1.76 \\
 198 & 2XMM  J120504.4+352209 &  2.278 &  2 &  -13.64 &  44.80 &  1.06 & 1.06 \\
 199 & 2XMM  J120943.4+393644 &  2.333 &  2 &  -13.53 &  44.94 & -0.79 & -0.79 \\
 200 & 2XMM  J121001.7+392151 &  2.955 &  3 &  -13.66 &  45.03 & -0.86 & 2.70 \\
 201 & 2XMM  J121111.1+393320 &  1.529 &  8 &  -13.68 &  44.37 & -0.81 & 2.93 \\
 202 & 2XMM  J121129.3+392513 &  1.671 &  8 &  -13.70 &  44.45 & -0.84 & 2.91 \\
 203 & 2XMM  J121426.5+140259 &  1.279 &  4 &  -12.83 &  45.05 & -0.04 & 3.02 \\
 204 & 2XMM  J121509.4+135450 & 0.8473 &  2 &  -12.76 &  44.70 &  2.78 & 2.78 \\
 205 & 2XMM  J121753.1+294305 &  1.647 &  8 &  -13.28 &  44.85 & -0.13 & 2.84 \\
 206 & 2XMM  J121808.5+471613 &  0.398 &  7 &  -12.75 &  43.95 &  1.48 & 3.19 \\
 207 & 2XMM  J121836.1+054628 & 0.7954 &  2 &  -13.08 &  44.32 &  3.09 & 3.09 \\
 208 & 2XMM  J121849.5+295451 &  0.962 &  8 &  -13.21 &  44.38 &  0.00 & 2.97 \\
 209 & 2XMM  J121911.1+470708 &  1.901 &  5 &  -13.60 &  44.66 &  1.71 & 2.87 \\
 210 & 2XMM  J121938.6+064022 &  1.187 &  2 &  -13.23 &  44.57 &  1.88 & 1.88 \\
 211 & 2XMM  J121952.2+472058 & 0.6531 &  6 &  -12.86 &  44.34 &  1.40 & 3.12 \\
 212 & 2XMMi J122051.4+282217 &  1.524 &  4 &  -13.58 &  44.47 & -0.20 & 0.37 \\
 213 & 2XMM  J122135.6+280614 &  3.288 &  5 &  -12.70 &  46.09 & -0.43 & 2.71 \\
 214 & 2XMM  J122222.7+041623 &   1.19 &  2 &  -13.54 &  44.27 &  2.53 & 2.53 \\
 215 & 2XMM  J122442.2+332941 & 0.7763 &  3 &  -13.02 &  44.35 &  2.09 & 2.49 \\
 216 & 2XMM  J122525.0+333651 & 0.7654 &  3 &  -13.21 &  44.14 &  2.09 & 2.49 \\
 217 & 2XMM  J122532.4+332532 & 0.5859 &  3 &  -12.87 &  44.22 &  2.14 & 2.54 \\
 218 & 2XMM  J122549.9+332455 &  1.133 &  2 &  -13.82 &  43.93 &  2.41 & 2.41 \\
 219 & 2XMM  J122556.1+130656 &   1.35 &  2 &  -13.48 &  44.45 &  2.59 & 2.59 \\
 220 & 2XMM  J122607.1+334559 &  1.158 &  3 &  -12.96 &  44.82 &  2.00 & 2.41 \\
 221 & 2XMM  J122627.0+332148 &  0.875 &  3 &  -13.67 &  43.83 &  2.07 & 2.47 \\
 222 & 2XMM  J122645.3+332801 &  3.339 &  2 &  -14.00 &  44.81 &  1.63 & 1.63 \\
 223 & 2XMM  J122703.3+125402 &  1.273 &  2 &  -13.60 &  44.27 &  2.60 & 2.60 \\
 224 & 2XMM  J122731.6+333259 &  1.608 &  2 &  -13.93 &  44.17 &  2.62 & 2.62 \\
 225 & 2XMM  J122923.7+075359 & 0.8538 &  2 &  -12.95 &  44.52 &  2.49 & 2.49 \\
 226 & 2XMM  J122931.2+015249 & 0.7704 & 19 &  -13.57 &  43.79 & -0.81 & 3.24 \\
 227 & 2XMM  J122934.7+015658 &  1.921 & 25 &  -13.29 &  44.99 & -1.43 & 3.03 \\
 228 & 2XMM  J122951.5+105827 &  1.847 &  2 &  -13.65 &  44.59 &  0.14 & 0.14 \\
 229 & 2XMM  J123035.4+153510 & 0.8028 &  2 &  -12.84 &  44.57 &  0.30 & 0.30 \\
 230 & 2XMM  J123049.7+640848 &  1.041 &  2 &  -13.57 &  44.10 &  2.72 & 2.72 \\
 231 & 2XMM  J123054.1+110011 & 0.2359 &  3 &  -11.74 &  44.45 &  0.50 & 2.86 \\
 232 & 2XMM  J123110.3+161258 &  1.453 &  2 &  -13.05 &  44.95 & -1.02 & -1.02 \\
 233 & 2XMM  J123126.4+105111 & 0.3039 &  3 &  -12.76 &  43.68 &  0.48 & 2.83 \\
 234 & 2XMM  J123147.1+123835 & 0.2916 &  2 &  -12.69 &  43.71 &  3.29 & 3.29 \\
 235 & 2XMM  J123148.0+143741 &  1.706 &  2 &  -13.37 &  44.80 &  1.90 & 1.90 \\
 236 & 2XMM  J123229.6+641115 & 0.7423 &  2 &  -12.98 &  44.34 &  2.79 & 2.79 \\
 237 & 2XMM  J123622.9+621526 &  2.588 &  7 &  -13.96 &  44.61 & -0.80 & 2.42 \\
 238 & 2XMM  J123716.0+620323 &  2.068 &  5 &  -14.31 &  44.04 &  0.24 & 2.49 \\
 239 & 2XMM  J123759.5+621102 & 0.9095 &  7 &  -13.05 &  44.48 & -0.52 & 2.69 \\
 240 & 2XMM  J123800.9+621336 & 0.4402 &  7 &  -13.21 &  43.60 & -0.40 & 2.81 \\
 241 & 2XMM  J123816.0+620208 &  1.005 &  3 &  -14.01 &  43.62 & -0.54 & 0.85 \\
 242 & 2XMM  J124126.5+323924 &  1.787 &  2 &  -13.50 &  44.70 &  0.95 & 0.95 \\
 243 & 2XMM  J124206.0+141920 &  1.951 &  2 &  -13.26 &  45.03 &  2.57 & 2.57 \\
 244 & 2XMM  J124207.6+333117 & 0.5148 &  2 &  -13.38 &  43.58 &  0.07 & 0.07 \\
 245 & 2XMM  J124300.3+113554 &   2.94 &  2 &  -13.12 &  45.57 &  2.81 & 2.81 \\
 246 & 2XMM  J124406.9+113524 &  1.344 &  2 &  -13.07 &  44.85 &  3.04 & 3.04 \\
 247 & 2XMM  J125317.6+310550 & 0.7824 &  2 &  -13.40 &  43.98 &  1.22 & 1.22 \\
 248 & 2XMM  J125344.9+305820 &  2.067 &  2 &  -13.42 &  44.93 &  0.98 & 0.98 \\
 249 & 2XMM  J125553.0+272405 & 0.3158 &  2 &  -12.22 &  44.25 &  2.92 & 2.92 \\
 250 & 2XMM  J125627.9+215406 &  1.871 &  5 &  -13.29 &  44.96 &  0.77 & 2.41 \\
 251 & 2XMM  J125629.6+271507 &  2.523 &  2 &  -13.56 &  44.98 &  2.49 & 2.49 \\
 252 & 2XMM  J125702.9+273801 &   1.13 &  3 &  -13.35 &  44.40 &  0.67 & 1.93 \\
 253 & 2XMM  J125708.4+271330 &  1.664 &  3 &  -13.27 &  44.86 &  0.15 & 2.91 \\
 254 & 2XMM  J125712.0+274216 & 0.7925 &  2 &  -13.03 &  44.37 &  2.01 & 2.01 \\
 255 & 2XMM  J125732.6+215708 &  1.934 &  4 &  -13.69 &  44.60 &  1.78 & 2.40 \\
 256 & 2XMMi J125732.9+473224 &  1.859 &  2 &  -13.57 &  44.67 & -0.18 & -0.18 \\
 257 & 2XMM  J125745.1+273210 &   1.56 &  5 &  -13.77 &  44.30 &  0.17 & 2.93 \\
 258 & 2XMM  J125803.0+345125 &  2.037 &  2 &  -14.01 &  44.33 &  0.93 & 0.93 \\
 259 & 2XMM  J125818.5+275937 &  1.723 &  2 &  -13.91 &  44.26 &  1.82 & 1.82 \\
 260 & 2XMM  J125831.7+275330 &   1.14 &  5 &  -13.04 &  44.72 & -0.82 & 2.93 \\
 261 & 2XMM  J125859.2+275308 &  1.135 & 12 &  -13.30 &  44.45 & -0.82 & 2.94 \\
 262 & 2XMM  J125903.9+344702 & 0.6075 &  3 &  -13.01 &  44.12 &  1.20 & 2.03 \\
 263 & 2XMM  J125923.3+272720 &   1.99 &  2 &  -13.46 &  44.85 &  0.15 & 0.15 \\
 264 & 2XMM  J125931.0+282706 &  1.094 &  7 &  -13.70 &  44.02 & -0.06 & 3.02 \\
 265 & 2XMM  J130002.7+345043 &  1.054 &  3 &  -13.25 &  44.43 &  1.10 & 1.92 \\
 266 & 2XMM  J130028.5+283010 & 0.6467 &  7 &  -12.09 &  45.10 &  0.05 & 2.95 \\
 267 & 2XMM  J130048.1+282321 &  1.923 &  7 &  -13.42 &  44.86 & -0.20 & 2.70 \\
 268 & 2XMM  J130100.8+281944 &   1.36 &  7 &  -13.09 &  44.85 & -0.11 & 2.79 \\
 269 & 2XMM  J130120.0+282137 &  1.369 &  8 &  -12.73 &  45.21 & -0.11 & 2.79 \\
 270 & 2XMMi J131134.9+231818 &  1.527 &  2 &  -13.50 &  44.55 &  0.52 & 0.52 \\
 271 & 2XMMi J131213.6+231958 &  1.514 &  3 &  -13.10 &  44.94 & -0.17 & 0.60 \\
 272 & 2XMMi J131236.2+231630 &  3.711 &  2 &  -13.86 &  45.04 &  0.24 & 0.24 \\
 273 & 2XMMi J131606.6+421513 &  1.841 &  2 &  -13.32 &  44.92 &  0.89 & 0.89 \\
 274 & 2XMMi J131712.9+420439 &  1.031 &  2 &  -12.78 &  44.88 &  1.03 & 1.03 \\
 275 & 2XMM  J132307.7+655446 & 0.6485 &  2 &  -12.91 &  44.28 &  2.64 & 2.64 \\
 276 & 2XMM  J132827.3+581839 &  3.139 &  4 &  -14.07 &  44.68 &  0.01 & 2.54 \\
 277 & 2XMM  J132938.5+471854 &  1.027 &  4 &  -13.27 &  44.39 &  0.31 & 2.78 \\
 278 & 2XMM  J133028.3+242253 &  1.919 &  2 &  -13.66 &  44.62 &  0.64 & 0.64 \\
 279 & 2XMM  J133114.5+241650 &  2.265 &  2 &  -13.72 &  44.71 &  0.59 & 0.59 \\
 280 & 2XMM  J133342.3+380336 &  1.077 &  3 &  -13.16 &  44.54 & -0.09 & 0.75 \\
 281 & 2XMM  J133417.5+375722 &  1.142 &  3 &  -13.25 &  44.51 & -0.10 & 0.74 \\
 282 & 2XMM  J133542.5+375542 &  1.899 &  3 &  -13.84 &  44.43 & -0.23 & 0.61 \\
 283 & 2XMM  J133807.5+242410 & 0.6313 &  4 &  -12.97 &  44.20 &  0.07 & 3.16 \\
 284 & 2XMM  J133859.2+272702 &  1.792 &  2 &  -13.62 &  44.58 &  1.17 & 1.17 \\
 285 & 2XMM  J133913.3+271818 & 0.6819 &  2 &  -12.71 &  44.53 &  1.39 & 1.39 \\
 286 & 2XMM  J133944.4-001451 &  1.269 &  2 &  -13.82 &  44.05 &  1.92 & 1.92 \\
 287 & 2XMMi J134050.7+301610 &  1.519 &  2 &  -13.52 &  44.53 & -0.10 & -0.10 \\
 288 & 2XMMi J134132.8+301326 & 0.7355 &  2 &  -13.20 &  44.12 &  0.06 & 0.06 \\
 289 & 2XMM  J134256.5+000057 & 0.8041 &  2 &  -12.76 &  44.65 &  2.48 & 2.48 \\
 290 & 2XMM  J134323.6+001223 & 0.8731 &  3 &  -12.79 &  44.70 &  1.93 & 2.47 \\
 291 & 2XMM  J134834.2+262205 & 0.9144 &  2 &  -12.67 &  44.86 &  2.69 & 2.69 \\
 292 & 2XMM  J134848.2+262219 & 0.5949 &  2 &  -12.93 &  44.17 &  2.77 & 2.77 \\
 293 & 2XMM  J134850.1+262503 &  2.915 &  2 &  -14.13 &  44.55 &  2.38 & 2.38 \\
 294 & 2XMM  J135038.6+601901 &  1.165 &  2 &  -13.49 &  44.29 &  1.98 & 1.98 \\
 295 & 2XMM  J135301.2+633256 &   3.16 &  2 &  -14.26 &  44.49 &  2.54 & 2.54 \\
 296 & 2XMM  J135418.1+635705 &  1.618 &  3 &  -13.78 &  44.33 & -0.11 & 2.74 \\
 297 & 2XMM  J135810.6+653740 &  1.112 &  3 &  -13.07 &  44.66 &  1.25 & 1.58 \\
 298 & 2XMM  J135842.7+652236 &  3.199 &  3 &  -13.81 &  44.95 &  0.95 & 1.28 \\
 299 & 2XMM  J140001.6-014924 &  1.754 &  3 &  -13.40 &  44.79 &  1.83 & 2.67 \\
 300 & 2XMM  J140146.5+024433 &  4.442 &  4 &  -13.69 &  45.39 &  0.26 & 2.73 \\
 301 & 2XMMi J140148.2-014514 &  1.795 &  2 &  -13.78 &  44.44 &  1.82 & 1.82 \\
 302 & 2XMM  J140349.4+432009 & 0.6664 &  2 &  -12.94 &  44.28 &  1.49 & 1.49 \\
 303 & 2XMM  J140354.6+543246 &  3.258 &  2 &  -14.07 &  44.71 &  2.26 & 2.26 \\
 304 & 2XMM  J140536.6+255140 & 0.9427 &  3 &  -13.01 &  44.55 &  0.04 & 2.96 \\
 305 & 2XMM  J140541.0+432537 & 0.5199 &  2 &  -13.23 &  43.74 &  1.53 & 1.53 \\
 306 & 2XMMi J140547.2+260629 & 0.7244 &  2 &  -13.60 &  43.71 &  0.09 & 0.09 \\
 307 & 2XMM  J140841.5+262943 &  1.885 &  2 &  -13.48 &  44.78 &  2.06 & 2.06 \\
 308 & 2XMM  J140949.0+261347 &  2.945 &  2 &  -13.94 &  44.75 &  1.93 & 1.93 \\
 309 & 2XMM  J141513.5+112216 &  1.554 &  2 &  -13.71 &  44.36 &  2.16 & 2.16 \\
 310 & 2XMM  J141515.8+112344 &  1.229 &  2 &  -13.48 &  44.36 &  2.22 & 2.22 \\
 311 & 2XMM  J141540.0+112407 &  1.074 &  2 &  -13.02 &  44.68 &  2.25 & 2.25 \\
 312 & 2XMM  J141546.2+112943 &   2.56 &  2 &  -13.74 &  44.82 &  2.02 & 2.02 \\
 313 & 2XMM  J141551.6+522743 &  2.587 &  2 &  -14.08 &  44.49 & -0.30 & -0.30 \\
 314 & 2XMM  J141642.3+521813 &  1.285 &  3 &  -13.68 &  44.20 & -0.98 & -0.11 \\
 315 & 2XMM  J141647.3+521115 &  2.152 &  3 &  -13.56 &  44.83 & -1.12 & -0.25 \\
 316 & 2XMM  J141652.0+113201 & 0.6881 &  2 &  -13.00 &  44.25 &  2.34 & 2.34 \\
 317 & 2XMM  J141745.6+250242 &   1.36 &  3 &  -13.61 &  44.32 & -0.00 & 1.93 \\
 318 & 2XMM  J141838.2+522400 &  1.118 &  3 &  -13.23 &  44.51 & -0.94 & -0.07 \\
 319 & 2XMMi J142258.2+193322 &  1.603 &  3 &  -12.65 &  45.45 & -0.12 & 0.19 \\
 320 & 2XMMi J142259.6+194458 &  1.129 &  3 &  -13.71 &  44.04 & -0.03 & 0.27 \\
 321 & 2XMM  J142325.4+384032 & 0.2489 &  2 &  -12.47 &  43.78 &  2.73 & 2.73 \\
 322 & 2XMM  J142335.9+383407 &  1.487 &  2 &  -13.60 &  44.42 &  2.43 & 2.43 \\
 323 & 2XMM  J142355.5+383150 &  1.207 &  2 &  -13.08 &  44.74 &  2.49 & 2.49 \\
 324 & 2XMM  J142406.6+383714 &  1.562 &  2 &  -13.11 &  44.96 &  2.42 & 2.42 \\
 325 & 2XMM  J142435.9+421030 &  2.218 &  2 &  -13.11 &  45.30 &  1.65 & 1.65 \\
 326 & 2XMM  J142455.5+421408 & 0.3162 &  2 &  -12.03 &  44.45 &  2.04 & 2.04 \\
 327 & 2XMM  J142519.0+422158 &  1.104 &  2 &  -13.45 &  44.28 &  1.83 & 1.83 \\
 328 & 2XMM  J142737.7+424450 &  1.953 &  7 &  -13.55 &  44.74 & -0.17 & 2.71 \\
 329 & 2XMM  J143025.8+415957 & 0.3524 &  3 &  -12.87 &  43.71 &  1.46 & 2.83 \\
 330 & 2XMM  J143440.4+484139 &  1.945 &  2 &  -13.50 &  44.79 &  1.01 & 1.01 \\
 331 & 2XMM  J143513.9+484149 &  1.887 &  2 &  -13.48 &  44.78 &  1.02 & 1.02 \\
 332 & 2XMM  J143621.2+484606 &  2.395 &  2 &  -13.61 &  44.88 &  0.95 & 0.95 \\
 333 & 2XMMi J143914.1+002320 & 0.8826 &  2 &  -13.05 &  44.45 &  1.99 & 1.99 \\
 334 & 2XMMi J143931.9+000453 &  1.405 &  2 &  -13.33 &  44.63 &  1.89 & 1.89 \\
 335 & 2XMMi J144008.6+001630 &  1.502 &  2 &  -13.08 &  44.95 &  1.87 & 1.87 \\
 336 & 2XMMi J144259.9-003725 &  1.817 &  2 &  -12.93 &  45.29 & -0.43 & -0.43 \\
 337 & 2XMMi J144305.1-004825 & 0.7007 &  2 &  -13.68 &  43.59 & -0.22 & -0.22 \\
 338 & 2XMMi J144308.1-004913 &  1.372 &  2 &  -12.97 &  44.97 & -0.36 & -0.36 \\
 339 & 2XMM  J144729.9+030520 &  1.782 &  2 &  -13.47 &  44.73 &  2.10 & 2.10 \\
 340 & 2XMM  J150424.9+102938 &  1.839 &  4 &  -12.31 &  45.92 & -1.17 & 2.76 \\
 341 & 2XMM  J150428.3+101856 &  1.011 &  4 &  -12.95 &  44.69 & -1.02 & 2.91 \\
 342 & 2XMM  J150545.6+014145 &  1.424 &  2 &  -13.22 &  44.76 &  1.94 & 1.94 \\
 343 & 2XMM  J150916.2+332730 &  1.656 &  2 &  -13.46 &  44.67 &  2.52 & 2.52 \\
 344 & 2XMM  J150948.6+333626 & 0.5124 &  2 &  -13.17 &  43.78 &  2.77 & 2.77 \\
 345 & 2XMM  J151126.4+565934 &  1.031 &  2 &  -13.45 &  44.21 &  2.12 & 2.12 \\
 346 & 2XMM  J151453.9+561032 &  1.287 &  2 &  -13.21 &  44.67 &  0.54 & 0.54 \\
 347 & 2XMM  J151510.1+562834 & 0.7207 &  2 &  -13.22 &  44.07 &  0.67 & 0.67 \\
 348 & 2XMM  J151651.2+562850 &  1.309 &  2 &  -13.35 &  44.55 &  0.54 & 0.54 \\
 349 & 2XMM  J152322.3+274931 &  0.424 &  2 &  -13.90 &  42.87 &  2.80 & 2.80 \\
 350 & 2XMM  J152553.8+513649 &  2.883 &  3 &  -12.83 &  45.84 &  0.03 & 1.54 \\
 351 & 2XMM  J153322.8+324351 &  1.899 &  3 &  -13.53 &  44.74 &  1.72 & 2.48 \\
 352 & 2XMMi J153434.8+574723 &  1.236 &  4 &  -13.53 &  44.31 & -0.06 & 0.42 \\
 353 & 2XMMi J153458.3+575625 &  1.129 &  4 &  -13.23 &  44.52 & -0.03 & 0.45 \\
 354 & 2XMM  J153617.2+544709 &   1.45 &  4 &  -13.15 &  44.85 & -0.08 & 2.68 \\
 355 & 2XMM  J153634.9+544317 & 0.9136 &  4 &  -13.23 &  44.30 &  0.02 & 2.79 \\
 356 & 2XMM  J153641.5+543505 &  0.447 &  4 &  -13.54 &  43.28 &  0.14 & 2.91 \\
 357 & 2XMMi J153716.2+574838 & 0.6406 &  4 &  -12.95 &  44.23 &  0.08 & 0.56 \\
 358 & 2XMM  J154234.3+540137 & 0.3959 &  2 &  -13.77 &  42.93 &  0.44 & 0.44 \\
 359 & 2XMM  J154316.4+540526 & 0.2452 &  2 &  -12.85 &  43.38 &  0.49 & 0.49 \\
 360 & 2XMM  J154359.4+535902 &  2.371 &  2 &  -13.05 &  45.43 &  0.06 & 0.06 \\
 361 & 2XMM  J154530.3+484608 & 0.3993 &  4 &  -12.16 &  44.55 &  0.15 & 3.06 \\
 362 & 2XMM  J154535.8+484713 &  1.404 &  3 &  -13.34 &  44.63 &  0.40 & 2.82 \\
 363 & 2XMM  J160318.0+430116 &  1.156 &  2 &  -13.42 &  44.36 & -0.04 & -0.04 \\
 364 & 2XMM  J160419.0+325631 &  2.281 &  2 &  -13.46 &  44.98 &  2.06 & 2.06 \\
 365 & 2XMM  J160501.3+174515 &  2.996 &  2 &  -13.75 &  44.95 & -0.36 & -0.36 \\
 366 & 2XMM  J160513.1+325829 &  2.256 &  5 &  -13.57 &  44.87 &  0.09 & 2.10 \\
 367 & 2XMM  J160603.6+174307 &  1.105 &  2 &  -13.58 &  44.15 & -0.08 & -0.08 \\
 368 & 2XMM  J160613.5+325554 &  1.874 &  6 &  -13.57 &  44.68 & -0.17 & 2.72 \\
 369 & 2XMM  J161706.8+122606 &  1.637 &  2 &  -13.36 &  44.76 & -1.12 & -1.12 \\
 370 & 2XMMi J162710.3+350118 &  2.288 &  2 &  -13.39 &  45.05 & -0.21 & -0.21 \\
 371 & 2XMMi J162722.4+351039 &  1.677 &  2 &  -13.31 &  44.83 & -0.12 & -0.12 \\
 372 & 2XMM  J162855.6+394034 &  1.521 &  3 &  -13.30 &  44.75 & -0.11 & 1.22 \\
 373 & 2XMM  J162937.1+394059 & 0.7241 &  3 &  -13.08 &  44.23 &  0.05 & 1.38 \\
 374 & 2XMM  J162940.4+393124 &  2.146 &  3 &  -13.50 &  44.89 & -0.21 & 1.12 \\
 375 & 2XMMi J163023.5+242546 &  2.312 &  3 &  -13.85 &  44.61 & -0.22 & 0.38 \\
 376 & 2XMM  J164056.2+363404 & 0.6761 &  2 &  -12.42 &  44.81 &  0.08 & 0.08 \\
 377 & 2XMM  J165430.7+395418 & 0.3397 &  2 &  -12.39 &  44.16 &  0.17 & 0.17 \\
 378 & 2XMM  J165713.2+352441 &  2.331 &  6 &  -13.86 &  44.61 & -0.22 & 2.52 \\
 379 & 2XMM  J170554.0+240638 & 0.9021 &  2 &  -13.27 &  44.25 &  2.29 & 2.29 \\
 380 & 2XMM  J170606.2+240305 & 0.7912 &  3 &  -13.40 &  43.99 &  1.94 & 2.31 \\
 381 & 2XMM  J170639.3+240606 & 0.8358 &  2 &  -13.44 &  44.01 &  1.93 & 1.93 \\
 382 & 2XMMi J171029.2+590833 & 0.8637 &  2 &  -12.87 &  44.61 &  0.87 & 0.87 \\
 383 & 2XMMi J171126.8+585543 & 0.5373 &  2 &  -12.77 &  44.23 &  0.96 & 0.96 \\
 384 & 2XMMi J171144.9+584917 &  1.533 &  2 &  -13.16 &  44.89 &  0.74 & 0.74 \\
 385 & 2XMM  J171359.4+640939 &   1.36 &  2 &  -13.52 &  44.42 &  1.29 & 1.29 \\
 386 & 2XMMi J171815.9+584613 &  1.413 &  6 &  -13.38 &  44.59 & -0.08 & 1.92 \\
 387 & 2XMMi J171818.1+584904 & 0.6344 &  7 &  -13.10 &  44.07 &  0.09 & 2.09 \\
 388 & 2XMMi J171930.2+584804 &  2.081 &  8 &  -13.05 &  45.30 & -0.19 & 1.82 \\
 389 & 2XMMi J172026.4+263816 &  1.145 &  3 &  -13.56 &  44.20 & -0.04 & 0.27 \\
 390 & 2XMMi J172052.1+590154 & 0.3512 &  8 &  -12.77 &  43.81 &  0.17 & 2.17 \\
 391 & 2XMMi J172130.9+584405 & 0.9997 &  7 &  -12.72 &  44.91 &  0.00 & 2.00 \\
 392 & 2XMM  J172255.3+320307 & 0.2752 &  3 &  -12.90 &  43.44 &  0.79 & 1.24 \\
 393 & 2XMM  J172256.7+321427 &  1.173 &  3 &  -12.83 &  44.95 &  0.56 & 1.00 \\
 394 & 2XMMi J172310.4+595105 & 0.9899 &  2 &  -13.33 &  44.29 &  0.47 & 0.47 \\
 395 & 2XMMi J172353.2+600002 &  1.453 &  2 &  -13.09 &  44.91 &  0.38 & 0.38 \\
 396 & 2XMM  J212912.1+120750 &  1.149 &  4 &  -13.50 &  44.27 &  1.90 & 2.64 \\
 397 & 2XMM  J215703.7-073829 &  1.899 &  2 &  -13.39 &  44.87 &  2.28 & 2.28 \\
 398 & 2XMM  J221640.1+001619 &  1.019 &  2 &  -13.47 &  44.18 &  1.14 & 1.14 \\
 399 & 2XMM  J221708.9+002718 &  1.112 &  2 &  -13.20 &  44.53 &  1.13 & 1.13 \\
 400 & 2XMM  J221715.1+002615 & 0.7532 &  2 &  -13.10 &  44.24 &  1.21 & 1.21 \\
 401 & 2XMM  J221738.4+001207 &  1.121 &  2 &  -13.76 &  43.99 &  1.12 & 1.12 \\
 402 & 2XMM  J221739.2+002903 &  1.646 &  2 &  -14.21 &  43.91 &  1.03 & 1.03 \\
 403 & 2XMM  J221751.3+001146 &  1.491 &  2 &  -13.64 &  44.39 &  1.05 & 1.05 \\
 404 & 2XMM  J221755.2+001513 &  2.092 &  2 &  -13.90 &  44.46 &  0.96 & 0.96 \\
 405 & 2XMM  J221806.6+000534 &  2.276 &  2 &  -13.95 &  44.49 &  0.93 & 0.93 \\
 406 & 2XMM  J231733.6+001129 & 0.8407 &  2 &  -13.29 &  44.16 &  1.98 & 1.98 \\
 407 & 2XMM  J231742.5+000535 & 0.3209 &  2 &  -12.26 &  44.23 &  2.13 & 2.13 \\
 408 & 2XMM  J231850.6+002554 &  1.592 &  2 &  -13.38 &  44.71 &  2.81 & 2.81 \\
 409 & 2XMMi J232810.5+150012 &  1.536 &  2 &  -13.64 &  44.42 &  2.33 & 2.33 \\
 410 & 2XMMi J235800.6-000107 &  1.454 &  2 &  -13.18 &  44.83 &  1.91 & 1.91 \\
 411 & 2XMMi J235844.9-000723 &  1.979 &  2 &  -13.52 &  44.79 &  1.82 & 1.82 \\
 412 & 2XMMi J235845.6-000458 &  1.609 &  2 &  -13.24 &  44.86 &  1.88 & 1.88 \\
\end{longtable}
}

\onltab{2}{
\begin{longtable}{rccrccl}
\caption{Swift sample.}\\
\hline\hline
$N_{sou}$ &name &$z$ &$N_{bin}$ &$\log f_X$ &$\log L_X$ & GRB field\\
 (1) &(2)  &(3)  &(4) &(5)  &(6)  &(7) \\
\hline
\endfirsthead
\caption{continued.}\\
\hline\hline
$N_{sou}$ &name &$z$ &$N_{bin}$ &$\log f_X$ &$\log L_X$ & GRB field\\
(1) &(2)  &(3)  &(4) &(5)  &(6)  &(7) \\
\hline
\endhead
\hline
\endfoot
 1 & SWIFTFT J005503+1408.0 & 1.67 & 14 & -13.17 & 45.39 & GRB050904  \\
 2 & SWIFTFT J020643+0023.7 & 1.22 & 15 & -13.60 & 44.85 & GRB060908 \\
 3 & SWIFTFT J020710+0010.3 & 0.92 & 16 & -13.62 & 44.37 & GRB060908 \\
 4 & SWIFTFT J020727+0028.9 & 1.18 & 12 & -13.53 & 44.65 & GRB060908 \\
 5 & SWIFTFT J075122+3109.8 & 1.31 & 24 & -13.44 & 44.77 & GRB070125 \\
 6 & SWIFTFT J082057+3153.9 & 1.07 & 13 & -12.72 & 45.26 & GRB051227  \\
 7 & SWIFTFT J084819+1336.1 & 1.48 & 22 & -13.02 & 45.35 & GRB051016B \\
 8 & SWIFTFT J092733+3022.7 & 1.34 &  8 & -13.42 & 45.04 & GRB050505  \\
 9 & SWIFTFT J092736+3020.0 & 1.26 & 12 & -12.94 & 45.28 & GRB050505  \\
10 & SWIFTFT J094821+3153.9 & 1.59 & 10 & -13.43 & 45.07 & GRB060108  \\
11 & SWIFTFT J101609+4336.2 & 0.59 & 21 & -12.80 & 44.59 & GRB050319  \\
12 & SWIFTFT J101727+4329.0 & 1.17 & 17 & -13.10 & 45.00 & GRB050319  \\
13 & SWIFTFT J113805+4047.5 & 2.18 &  8 & -13.13 & 45.62 & GRB050215B \\
14 & SWIFTFT J114502+5957.3 & 1.64 & 75 & -13.39 & 45.06 & GRB060319  \\
15 & SWIFTFT J120215+1045.3 & 1.32 & 29 & -13.11 & 45.07 & GRB050408  \\
16 & SWIFTFT J121017+3956.7 & 0.40 & 15 & -12.74 & 44.20 & GRB070419A \\
17 & SWIFTFT J121645+3529.6 & 2.01 & 21 & -13.39 & 45.33 & GRB060712  \\
18 & SWIFTFT J124958+3028.8 & 1.63 &  5 & -13.13 & 45.28 & GRB050520  \\
19 & SWIFTFT J131524+1638.0 & 1.44 & 19 & -13.64 & 44.66 & GRB070406  \\
20 & SWIFTFT J133128+4209.7 & 0.94 & 11 & -13.67 & 44.11 & GRB051008 \\
21 & SWIFTFT J140704+2735.8 & 2.22 & 17 & -13.22 & 45.55 & GRB060204B \\ 
22 & SWIFTFT J141221+1657.9 & 1.87 &  7 & -13.26 & 45.32 & GRB060801 \\
23 & SWIFTFT J143646+2745.0 & 0.22 & 19 & -13.50 & 42.43 & GRB050802 \\
24 & SWIFTFT J144339+1229.4 & 1.98 &  5 & -13.27 & 45.31 & GRB060805 \\
25 & SWIFTFT J144419+1236.3 & 1.51 &  6 & -13.49 & 44.95 & GRB060805  \\
26 & SWIFTFT J160757+3221.8 & 1.42 & 12 & -13.52 & 44.81 & GRB060219  \\
27 & SWIFTFT J165004+3133.9 & 1.70 & 11 & -13.50 & 44.98 & GRB060807  \\
\end{longtable}
}

\end{document}